\documentclass{aa}  
\usepackage{graphicx}
\usepackage[varg]{txfonts}
\usepackage{pdflscape}
\usepackage[font=small, labelfont=bf,justification=justified,format=plain]{caption}
\usepackage{natbib,twoopt}
\usepackage{amsmath}
\usepackage{moresize}
\usepackage{afterpage}
\bibpunct{(}{)}{;}{a}{}{,} 
\usepackage{placeins}
\usepackage{float}
\usepackage{booktabs} 
\usepackage{amsmath}
\usepackage[raggedright]{titlesec}
\usepackage{longtable}
\usepackage{url}
\usepackage{xcolor}
\usepackage{subcaption}
\usepackage{commath}
\usepackage{babel}
\usepackage[utf8]{inputenc}
\usepackage[T1]{fontenc}     
\usepackage[autostyle=true]{csquotes} 
\usepackage{siunitx}
\usepackage{IEEEtrantools}                       
\usepackage{comment}
\usepackage{dirtytalk}
\usepackage{multirow}
\usepackage{multicol}
\usepackage[colorlinks=true, linkcolor=blue, citecolor=blue, urlcolor=blue]{hyperref}

\begin{document} 

   \title{Massive star formation at the Galactic crossroads: Insights from G358.69+0.03 in the Galactic center}
   \titlerunning{Massive star formation in G358.69+0.03}
   \author{A. Cheema\inst{1}\fnmsep\thanks{Member of the International Max Planck Research School (IMPRS) for Astronomy and Astrophysics at the Universities of Bonn and Cologne.} \and V. S. Veena\inst{1,2} \and K. M. Menten\inst{1}\fnmsep\thanks{Deceased} \and T. S. Pillai\inst{1,3} \and S. A. Dzib\inst{1} \and A. Brunthaler\inst{1} \and S. Khan\inst{1} \and R. Dokara\inst{1} \and M. R. Rugel\inst{4,5} \and Y. Gong\inst{1,6}
    }

   \institute{Max Planck Institute for Radioastronomy (MPIfR), Auf dem H\"ugel 69, 53121 Bonn, Germany \\
   \email{acheema@mpifr-bonn.mpg.de}
         \and 
       Department of Astronomy \& Astrophysics, Tata Institute of Fundamental Research,
       Homi Bhabha Road, Mumbai 400005, India
        \and
        MIT Haystack Observatory, 99 Millstone Road , Westford, MA 01827, USA
        \and
        Deutsches Zentrum f\"{u}r Astrophysik, Postplatz 1, 02826 G\"{o}rlitz, Germany
        \and
        National Radio Astronomy Observatory, P.O. Box O, 1003 Lopezville Road, Socorro, NM 87801, USA
        \and
        Purple Mountain Observatory, and Key Laboratory of Radio Astronomy, Chinese Academy of Sciences, 10 Yuanhua Road, Nanjing 210023, PR China        
             }

   \date{Received ; accepted }

    \abstract
    {We investigated the high-mass star formation activity in a subregion of the Sagittarius\,E star-forming complex, centered at (\textit{l}, \textit{b}) = (358.69\textdegree, 0.03\textdegree), where infrared and radio sources trace a prominent U-shaped structure that has not been identified in previous studies. We used radio continuum data from the Global View on Star Formation (GLOSTAR) survey, which is a wide-band radio (4--8 GHz) survey of the Milky Way that combines data from the \textit{Karl G. Jansky} Very Large Array and the Effelsberg 100 m telescope. Using \texttt{BLOBCAT} source extraction software, we identified 49 compact radio sources. Based on multiwavelength associations and spectral index estimates, we identified GLOSTAR counterparts to 27 previously confirmed \ion{H}{ii} regions, detected radio emission from 3 WISE ``radio-quiet'' candidates, and report 5 new \ion{H}{ii} region candidates. The derived physical properties indicate that most are relatively evolved \ion{H}{ii} regions. We find around 50 cold dust clumps, predominantly toward the south and southeast. Mid-infrared flux-ratio maps ([4.5]/[3.6]) show localized shock enhancements along the arc and adjacent clumps, and 15 clumps exhibit SiO emission with broad components indicative of shocks. Together with CO data, the SiO velocity components delineate a continuous ($>$100~km\,s$^{-1}$) velocity bridge that links the far dust-lane inflow to the central molecular zone (CMZ) stream. The largest concentration of clumps and compact \ion{H}{ii} regions lies at this interface. These combined diagnostics favor a scenario in which bar-driven cloud–cloud collision at the far dust-lane--CMZ interface compressed the gas and triggered the observed high-mass star formation.}


    \maketitle
    \nolinenumbers

    \section{Introduction}

    Massive star formation and its radiative and mechanical feedback are critical drivers of galactic evolution. In this context, the nuclear regions of galaxies are particularly important, as they often exhibit more extreme physical, chemical, and kinematic conditions than the galactic disks. However, studying extragalactic nuclei poses significant observational challenges due to high interstellar extinction and the need for subarcsecond angular resolution to overcome severe source crowding. The Milky Way’s Galactic center (GC) offers a unique local laboratory for investigating massive star formation under such extreme conditions \citep[and references therein]{2012Shetty, 2013Longmore, 2016Ginsburg, 2017Barnes, 2016Henshaw, 2022Henshaw}. By exploring the physical processes and chemistry of the Milky Way’s nucleus, we not only gain insights into its formation and evolution but also establish a critical template for interpreting the central regions of nearby galaxies. \\

    \begin{figure*}[!htbp]
        \centering
        \includegraphics[width=\textwidth]{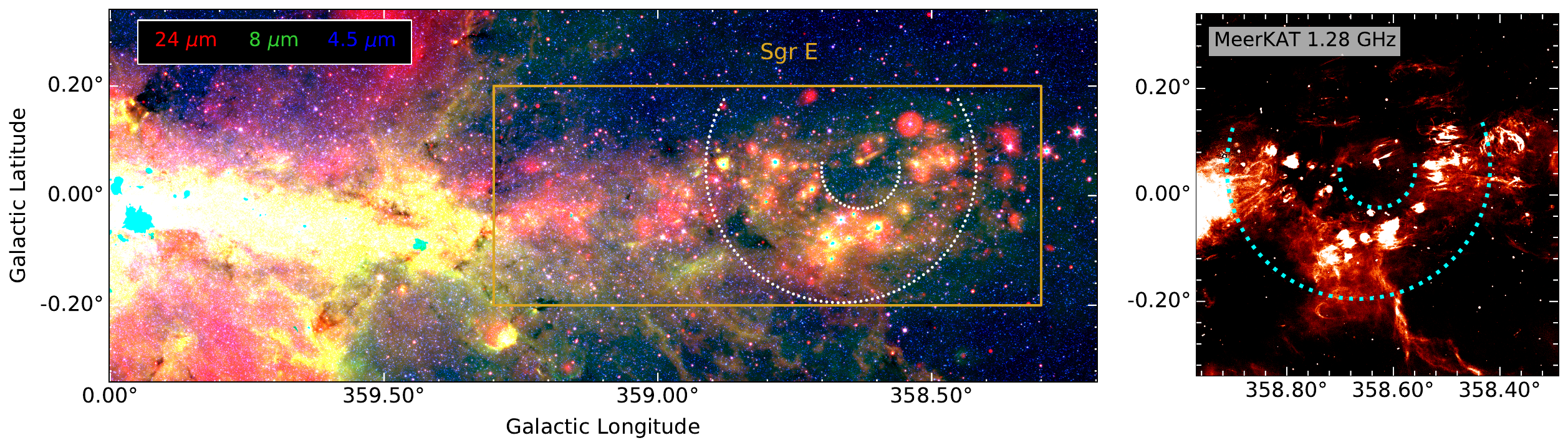}
        \caption{\textit{Left}: Three-color infrared view of the Galactic plane on the negative longitude side, with MIPSGAL 24 $\mu$m emission in red, GLIMPSE 8 $\mu$m in green, and GLIMPSE 4.5 $\mu$m in blue. The golden box shows the Sgr\,E star-forming complex \citep[358.3\textdegree<\textit{l}<359.3\textdegree, \ |\textit{b}| < 0.2\textdegree;][]{2020Anderson}. \textit{Right}: 1.28 GHz MeerKAT image of the target region \citep{2022meerkat}. The dashed partial circles in both panels highlight the arc-like or U-shaped morphology of G358.69+0.03.}
        \label{mainrgb}
    \end{figure*}  

    The central molecular zone (CMZ), which is the innermost few hundred parsecs of the Milky Way, holds approximately 2-6 $\times$ 10$^7$ \(M_\odot\) of molecular material, accounting for $\lesssim$10$\%$ of the total molecular gas and $\thicksim$80$\%$ of  the dense gas within the Milky Way \citep[etc.]{2007Ferriere, 2014Molinari, 2016Roman-duval, 2025Battersby}. It hosts plentiful energetic phenomena, most importantly the ongoing and past massive star formation activity in complexes such as Sgr\,B2 and Sgr\,C. The CMZ roughly spans the Galactic longitude range of -1\textdegree \ $\lesssim$ \textit{l} $\lesssim$ 1.7\textdegree \ \citep{2022Henshaw}, with mass inflow into the region facilitated by two dust lanes along the Galactic bar. The near-side dust lane, called the connecting arm, is situated at \textit{l} > 0\textdegree, and the far-side dust lane merges with the CMZ at  \textit{l} < 0\textdegree \ \citep{2008Marshall, 2019Sormani, 2022Henshaw}. Due to the energetic conditions, the CMZ is expected to have a high star formation rate (SFR). 
    
    \citet{2009Yusefzadeh} studied the central 400 × 50 pc region of the GC and observed an asymmetry in the distribution of the candidate young stellar objects (YSOs), with a higher concentration toward the negative Galactic longitudes. They estimated the SFR based on the identification of Stage I YSO candidates to be about 0.14 \(M_\odot\) yr$^{-1}$, which is much higher than the SFR estimates of the CMZ in literature of, for example, $\thicksim$ 0.08  \(M_\odot\) yr$^{-1}$ \citep{2024Hatchfield} and $\thicksim$ 0.068 \(M_\odot\) yr$^{-1}$ \citep{2021Hans}. The asymmetry of YSOs has been a matter of scientific controversy, with many studies providing conflicting conclusions. For example, \citet{2015Koepferl} studied the sample of YSOs from \citet{2009Yusefzadeh} using radiative transfer models and showed that embedded main sequence stars can masquerade as YSOs. This misclassification could lead to incorrect conclusions regarding the distribution of YSOs and, consequently, affect the inferred SFR of the CMZ. 
   
    \citet{2020Anderson} further investigated massive star formation activity along the negative Galactic longitudes, in particular in the star-forming complex Sagittarius E \citep[Sgr\,E; 358.3\textdegree < \textit{l} < 359.3\textdegree; |\textit{b}|<0.2\textdegree;][]{1992Liszt, 1993Gray, 1994Gray, 1996Cram}. Sgr\,E is located at a projected distance of $\thicksim$ 220 pc from the GC, at the intersection of the CMZ and the far dust lane. Using 1.28 GHz MeerKAT GC data and radio recombination line (RRL) observations from the Green Bank Telescope, they found 19 \ion{H}{ii} regions and 43 \ion{H}{ii} region candidates embedded in a molecular cloud of mass $3\times10^5 \ M_\odot$, highlighting the region as a hot spot of high-mass star formation in the GC. They propose that Sgr\,E formed upstream in the far dust lane a few million years ago, and would likely overshoot the CMZ and crash into the near dust lane.

\begin{figure*}[]
        \centering
        \includegraphics[height = 0.33 \textwidth]{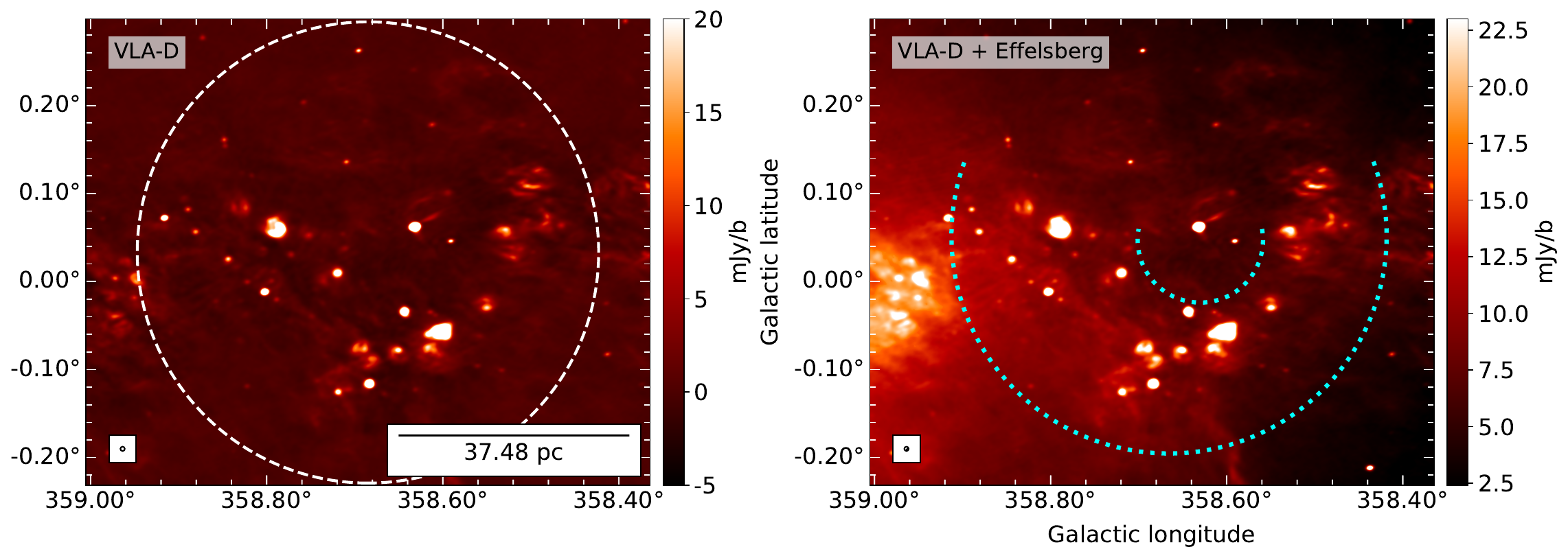}
        \caption{\textit{Left}: GLOSTAR D configuration radio continuum averaged intensity map at 5.8 GHz with an angular resolution of $\thicksim$ 18$''$. The dashed white circle marks the target region. \textit{Right}: GLOSTAR combination image (VLA-D+Effelsberg). The dashed cyan partial circles highlight the arc-like or U-shaped morphology of G358.69+0.03. The beam sizes are shown in the lower-left corner of each panel.}
        \label{target}
    \end{figure*}

    Upon further investigation, we noticed a striking arc-like or U-shaped distribution of the compact \ion{H}{ii} regions in a subregion of the Sgr\,E complex (see Fig.~\ref{mainrgb}). The region also shows relatively strong radio and infrared emission along a similar U-shaped morphology. In the present study, we conducted an extensive analysis of the star formation activity within this subregion of the Sgr\,E complex, hereafter referred to as G358.69+0.03 (see Fig.~\ref{target}). It has a radius of approximately 0.26\textdegree, which corresponds to about 37.48 pc at the GC distance of 8.2 kpc \citep{2019Abuter}, and is centered at Galactic coordinates \textit{l} = 358.69\textdegree, \ \textit{b} = 0.03\textdegree. We further refined the population of \ion{H}{ii} regions, tracers of massive star formation activity, using observations from the \textit{Karl G. Jansky} Very Large Array (VLA) D configuration at a higher frequency of 5.8 GHz. Additionally, we examined the distribution of cold dust clumps and the signature of shocks, and explored the possibility of a triggered star formation scenario to explain the region's morphology.
    
    Section \ref{secdata} gives a summary of the data used in this study. Section \ref{sec3} describes the source extraction procedures and the multiwavelength analysis performed to identify the candidate \ion{H}{ii} (c\ion{H}{ii}) regions. Section \ref{results} provides details on the properties of the target region and the distribution of cold dust clumps and shock signatures across G358.69+0.03. Section \ref{discussion} discusses the triggered star formation scenario invoked to explain the morphology of our target region. We present our conclusions in Sect.~\ref{conlusions}.

    \section{Data}\label{secdata}
    \subsection{Radio continuum observations}
    Over the past few decades, there have been a large number of Galactic plane surveys of the Milky Way, observing both spectral-line and continuum emission covering the entire wavelength range from the near-infrared (NIR) to the radio \citep[etc.]{1994Becker, 2008ukidds, 2009Carey, 2009Schuller, 2011Aguirre, 2016Molinari, 2016Beuther}. To study high-mass star formation activity at the negative Galactic longitudes of the Milky Way, we used data from the Global View on Star Formation (GLOSTAR\footnote{\url{https://glostar.mpifr-bonn.mpg.de/glostar/}}) survey \citep{2019Sac, 2021Brunthaler}. \\
    
    The GLOSTAR survey is the most sensitive radio survey ($\thicksim$ 60--150 $\mu$Jy beam$^{-1}$) of the Galactic plane at the \textit{C}-band of 4--8 GHz.  The observations were conducted using the upgraded VLA of the National Radio Astronomy Observatory (NRAO) and the Effelsberg 100 m telescope with increased frequency coverage and unprecedented sensitivity. The main objective of the survey is to study massive star formation activity in the Galactic mid-plane of the Milky Way. The survey simultaneously observes radio continuum in full polarization and spectral line emission such as the 6.7 GHz methanol maser line, 4.8 GHz formaldehyde line and seven RRLs covering the velocity range of --150 km\,s$^{-1}$ to +150 km\,s$^{-1}$ \citep{2021Brunthaler}.  In the continuum mode, the survey observes different evolutionary stages of the tell-tale tracers of high-mass star formation, namely hypercompact, ultracompact, and compact \ion{H}{ii} regions \citep{2019Sac, 2021Hans, 2023Sergio, 2023yang, 2024sarwar, 2024sac}.

    The GC (|\textit{l}| < 2\textdegree; |\textit{b}| < 1\textdegree) was observed using the \textit{C}-band receivers (4--8 GHz) of the VLA in the DnC configuration, i.e., the northern arm of the VLA was in the C configuration and the other two arms were in the D configuration, providing an angular resolution of 18$''$. This configuration gives a more circular beam for low-declination sources. The correlator setup consists of two 1 GHz sub-bands, centered at 4.7 and 6.9 GHz; they have primary beams with full widths at half maximum (FWHMs) of $\thicksim$9$'$ and $\thicksim$6$'$, respectively \citep{2019Sac, 2021Brunthaler}. Each sub-band consisted of 8 spectral windows of 128 MHz each, which are further divided into 64 channels of 2 MHz each. The GC observations were conducted in September-October 2014 and January 2016. The data were then calibrated using a modified pipeline created using the \textit{Obit} software package \citep{2008Cotton}, which is explained in more detail in \citet{2019Sac}. The resultant VLA-D integrated intensity radio continuum map with an effective frequency of 5.8 GHz and a pixel size of 2.5$''$ is shown in the left panel of Fig.~\ref{target}. 
    
    The VLA data were complemented with low-resolution $\thicksim$ 150$''$ images from single-dish observations of the Effelsberg 100 m radio telescope. This was done to provide the zero spacings for the D configuration data to recover the extended emission. The image combination process and the caveats are discussed in detail in \cite{2023Rohit}. The single dish Effelsberg observations were carried out in July 2020 and April 2021. The resultant integrated intensity VLA-D+Effelsberg map of the target region is shown in the right panel of Fig.~\ref{target}.

    \subsection{APEX nFLASH follow-up}
    To study SiO emission from  a cluster of dust clumps located toward the south of the target region, we selected 17 sources from the APEX Telescope Large Area Survey of the Galaxy (ATLASGAL) Compact Source Catalogue \citep[CSC;][]{2014urquhart}. We carried out the standard position switch observations using the Atacama Pathfinder Experiment (APEX) 12m telescope \citep{2006gusten} in March 2023 (ID: M-0111.F-9523C-2023). The total observing time was 11.8 hours. We used the nFLASH230 receiver as the frontend instrument tuned at 230.79  GHz in the upper sideband (USB) and the fast Fourier transform spectrometer (FFTS1) as the backend instrument. The spectral resolution is 0.25 km\,s$^{-1}$, and the APEX/nFLASH230 beam FWHM is $\thicksim$26.2$''$\footnote{\url{https://www.apex-telescope.org/telescope/efficiency/}}. The receiver covers a total of 32 GHz IF instantaneous bandwidth, including both sidebands and polarizations. The setup covers the SiO (5 -- 4) transition with a rest frequency of 217104.984 MHz, which is considered a good shock tracer \citep{1997Schilke}.  
    
    The antenna temperatures ($T_A^*$) were converted to the main beam brightness temperature ($T_{\rm MB}$) scale using a beam efficiency factor of 0.83. The data reduction was performed using the GILDAS\footnote{\url{http://www.iram.fr/IRAMFR/GILDAS}} software \citep{2005pety}. 

    \subsection{Archival data}\label{archival-data}
    To carry out a comprehensive multiwavelength analysis, we used archival data from complementary multiwavelength surveys.

    \subsubsection{\textit{Spitzer} Space Telescope}
    We used the images from the Galactic Legacy Infrared Mid--Plane Survey Extraordinaire (GLIMPSE), a legacy program observed using the Infrared Array Camera (IRAC) on board the \textit{Spitzer} Space Telescope  \citep{2003Benjamin,2005Churchwell} to study the mid-infrared (MIR) morphology of the target region. It covers the wavelength range from 3.6~$\mu$m to 8.0~$\mu$m with an angular resolution of $\thicksim$ 2$''$. We also used the 24 $\mu$m MIR image with an angular resolution of 6$''$ covered using the Multiband Infrared Photometer for \textit{Spitzer} (MIPSGAL) on board the \textit{Spitzer} Space Telescope \citep{2009Carey}.  We also utilized the GLIMPSE II Epoch 1 Catalog \citep{2020ipac.data.I227G} and the MIPSGAL 24 $\mu$m Catalog \citep{2020mipsgal} obtained from the NASA/IPAC Infrared Science Archive (IRSA\footnote{\url{https://irsa.ipac.caltech.edu/frontpage/}}). 
    
    \subsubsection{WISE}
    We utilized the Wide-field Infrared Survey Explorer (WISE) maps at wavelengths of 3.4, 4.6, 12, and 22 $\mu$m, having angular resolutions of 6.1$''$, 6.4$''$, 6.5$''$, and 12.0$''$, respectively \citep{2010Wright}, to study the MIR signature of the target region. We also utilized the WISE All-sky Source Catalog \citep{2020Wise} accessed from IRSA.

    \subsubsection{\textit{Herschel} Hi--GAL survey}
    To investigate the structure associated with the cold and diffuse interstellar dust in the target region, we used the images and photometric compact source catalogs from the \textit{Herschel} infrared Galactic Plane Survey (Hi-GAL), which observes the inner Galactic plane (68\textdegree \ $\gtrsim$ \textit{l} $\gtrsim$ --70\textdegree, |\textit{b}| $\leq$ 1\textdegree) of the Milky Way at the far-infrared (FIR) wavelengths of 70, 160, 250, 350, and 500 $\mu$m using Photodetector Array Camera and Spectrometer (PACS) for the first two and Spectral and Photometric Imaging Receiver (SPIRE) for the last three \citep{2016Molinari}. The angular resolutions are 6$''$, 12$''$, 18$''$, 24$'',$ and 35$'',$ respectively. We also used the 360\textdegree \ CSC \citep{2021Elia} to study the properties of Hi-GAL clumps.

    \subsubsection{APEX ATLASGAL survey}
    We used the images from ATLASGAL taken using the Large APEX Bolometer Camera (LABOCA) at the submillimeter wavelength of 870 $\mu$m with an angular resolution of 19.2$''$ \citep{2009Schuller}. The primary goal is to study the earliest embedded stages of massive star formation activity (pre- and proto-stellar clumps) using the optically thin, thermal emission from dust at submillimeter wavelengths. We also used the ATLASGAL CSC \citep[][]{2014urquhart} to study the distribution of cold gas clumps. 

    \subsubsection{CHIMPS2}
    To explore the large-scale dynamics of the G358.69+0.03 region, we used $^{12}$CO($J=3-2$) data from the CO Heterodyne Inner Milky Way Plane Survey 2 \citep[CHIMPS2;][]{2020chimps2}, obtained with the \textit{James Clerk Maxwell} Telescope (JCMT). The spatial resolution, spectral resolution, and root mean square (rms) noise are 15$''$, 1.0 km\,s$^{-1}$, and 0.8 K, respectively. The details of the observations and data reduction are provided in \citet{2020chimps2}, and the reduced data are available for download from the Canadian Advanced Network for Astronomical Research (CANFAR) archive\footnote{\url{https://www.canfar.net/citation/landing?doi=20.0004}}.
    
    \section{Radio emission and \ion{H}{ii} region population in G358.69+0.03 }\label{sec3}

    A multiwavelength view of G358.69+0.03 is presented in Fig.~\ref{mainrgb}. The infrared three-color composite maps reveal several compact sources embedded within diffuse 8~$\mu$m emission, spanning a region of $\thicksim$30 arcminutes. To characterize the nature of these compact sources and the associated diffuse emission, we analyzed 5.8 GHz data from the GLOSTAR VLA-D configuration and the VLA+Effelsberg observations, shown in Fig.~\ref{target}. The VLA image reveals numerous compact sources arranged in an arc-like or U-shaped distribution, accompanied by faint extended emission. The VLA+Effelsberg image recovers the extended emission more effectively, highlighting a shell-like structure with compact sources distributed around its periphery. To investigate the population of these compact radio sources, we compiled a catalog, which is detailed in the following subsections.
    
    \subsection{Source extraction}
    We focused on compact \ion{H}{ii} regions and other emission sources of diameter greater than $\thicksim$0.7 pc, which can be traced well with GLOSTAR VLA-D configuration images at a distance of 8.2 kpc. The source extraction was performed using this VLA-D configuration image, following the procedures described by \citet{2019Sac} and \citet{2023Sergio}.

    We used \texttt{BLOBCAT\footnote{\url{https://blobcat.sourceforge.net/}}}, a \texttt{Python} based source extraction software that uses the flood fill algorithm to detect and catalog ``blobs'' or island of pixels representing sources in 2D astronomical images \citep[FITS files;][]{2012blobcat}. The software handles both Gaussian and non-Gaussian sources and produces an output catalog listing the properties of the extracted blobs. \texttt{BLOBCAT} requires two input FITS images and some input arguments to perform the source extraction. The first is the integrated intensity map, i.e., the VLA-D configuration image with an effective frequency of $\nu$ $\approx$ 5.8 GHz and a spatial resolution of 18$''$. The second is the background rms noise image, which we created using the \texttt{SExtractor\footnote{\url{http://star-www.dur.ac.uk/~pdraper/gaia/gaia.htx/extract.html}}} tool \citep{1996Bertin} from the Graphical Astronomy and Image Analysis (GAIA) tool package. \texttt{SExtractor} estimates the noise level at each pixel by extracting the pixel values within a local mesh of user-specified size, and iteratively clipping the extreme values until convergence is reached at a chosen $\sigma$ value about the median \citep{2012blobcat}. Following \citet{2019Sac}, we used a detection threshold of 5$\sigma$, a minimum size of 5 pixels, and mesh size of 80 $\times$ 80 pixels$^2$ to create the noise map, which represents the position-dependent noise excluding most of the real emission from the sources. The rms noise toward the target region was estimated using the Common Astronomy Software Applications package \citep[CASA;][]{2007McMullin, 2022Casa}, as $\sigma_{\mathrm{rms}}$ $\approx$ 356 $\mu$Jy beam$^{-1}$. 

    \texttt{BLOBCAT} creates a signal-to-noise ratio (S/N) map by taking pixel-by-pixel ratio of the input integrated intensity and noise maps and extracts blobs using the additional user-specified arguments. \citet{2013Purcell} noted that in their radio continuum images of the Galactic plane observed at 5 GHz using the Coordinated Radio and Infrared Survey for High-Mass Star Formation \citep[CORNISH;][]{2012cornish}, a majority of radio detections with an S/N of below 4.5 were spurious sources. Hence, we used a S/N threshold for detecting the blob of $T_d$ = 5 (\texttt{dSNR}). The minimum accepted blob size in pixels is taken as, \texttt{minpix} = 20, which is the number of expected pixels of size 2.5$''$ covering 50\% of the beam area of size (circular beam of diameter) 18$''$. We specified a cutoff S/N threshold for flooding of $T_f$ = 3 (\texttt{fSNR}). We tried a number of threshold limits and chose this value based on the visual inspection and comparison with other wavelength images such that \texttt{BLOBCAT} does not misidentify distinct radio sources in crowded fields as one single source. However, in two specific regions -- G358.51+0.05 and G358.69-0.08, with sizes of 288$''$ and 180$'',$ respectively -- multiple distinct sources were incorrectly identified as one. So, we ran \texttt{BLOBCAT} separately on these regions with higher cutoff S/N thresholds of 4.6 and 7.6, respectively (see Fig.~\ref{morphology}, for example). A total of 49 sources were extracted, each with a flux above five times the local noise and a size greater than half the beam size. Finally, we performed an individual visual inspection of all these sources and did not find any obvious image artifacts, such as sidelobes from very bright sources. The most reliable sources have peak flux values above 7$\sigma_{\mathrm{noise}}$. We have a total of 39 such sources as well as 10 sources with peak fluxes between 5 and 7 times the local noise. Figure~\ref{blobcat} shows all the \texttt{BLOBCAT}-extracted sources toward our target region.    
    \subsection{Multiwavelength associations}
    To better understand the nature of these 49 radio sources, we searched for their counterparts at other wavelengths. We used a number of infrared surveys that cover the Galactic plane, including the ATLASGAL submillimeter survey \citep{2009Schuller} and the Multi-Array Galactic Plane Imaging Survey (MAGPIS) radio survey \citep{2005White}; the full list is provided in Sect. \ref{archival-data}. The catalogs were selected such that their spatial resolutions and/or wavelengths are comparable with those of the GLOSTAR survey. Our main focus is on identifying the c\ion{H}{ii} regions, which are the most unambiguous tracers of massive star formation activity \citep{2014Anderson}. \ion{H}{ii} regions are the brightest objects in the Galactic plane at infrared and radio wavelengths \citep{2014Anderson}. They also have characteristic 8 $\mu$m emission shells that trace the warm polycyclic aromatic hydrocarbons (PAHs), surrounding emission from the warm dust traced either by 24~$\mu$m or radio emission \citep{2012Simpson}. At MIR wavelengths, \ion{H}{ii} regions also show a characteristic morphology of $\thicksim$ 20~$\mu$m emission coincident with radio emission tracing the ionized gas, which is surrounded by $\thicksim$ 10~$\mu$m emission \citep[and references therein]{2014Anderson}. The FIR and submillimeter emission associated with cold dense gas and dust can provide evidence of molecular clumps with embedded \ion{H}{ii} regions \citep{2013Urquhart}. \ion{H}{ii} regions can show weak signature in the NIR during the deeply embedded stage in the molecular clouds \citep{2021Yang}. 
    
    The NIR counterparts were identified using the Two Micron All Sky Survey (2MASS) All-Sky Point Source Catalog \citep[PSC;][]{2006Skrutskie, 2003Cutri}. Within a 4$''$ search radius, we found 31 cross-matches. Furthermore, we performed a visual inspection to identify multiwavelength counterparts for the \texttt{BLOBCAT}-extracted radio sources across various images from 3.6 $\mu$m (MIR) to 870 $\mu$m (submillimeter), as well as the radio 1.28 GHz MeerKAT GC image \citep{2022meerkat}. Table~\ref{count} marks the counterparts found for each of the 49 extracted sources across different infrared and radio wavelengths. Additionally, we searched for counterparts in various catalogs listed in detail in Table  \ref{tab:numberofcounterparts}, which also includes the respective angular resolution and the search radius used to identify potential cross-matches. Notably, some sources had potential matches within the specified search radius but lacked a clear visual signature in the respective maps. These sources are marked with the superscript ``\textit{c}'' in Table~\ref{count}.

    All the \texttt{BLOBCAT}-extracted radio sources have a counterpart in at least one other published radio or infrared catalog, which eliminates the possibility of false detection. We found the most cross-matches in the MIR WISE All-Sky Source Catalog and the least in the ATLASGAL dust continuum and Hi-GAL infrared catalogs. All sources except G358.747--0.154, G358.622+0.093, G358.557+0.243, G358.633+0.240, and G358.688+0.182 have a counterpart in the MeerKAT GC image. 
    
    \subsection{Spectral index}\label{sec:si}
    The spectral index, $\alpha$ ($S_\nu \ \propto \nu^\alpha$, where $S_\nu$ is the flux density at the observed frequency, $\nu$) helps distinguish between thermal and nonthermal radio sources. \ion{H}{ii} regions emit thermal bremsstrahlung radiation and usually have spectral indices of around --0.1 (optically thin) and 2 (optically thick) at high and low frequencies, respectively.
    
    The GLOSTAR-VLA image products consist of nine sub-images in the 4-8 GHz frequency range, which we used to estimate the in-band spectral index. Following the procedure explained by \citet{2024sac}, we estimated the spectral indices of compact sources, defined as those with a ratio of integrated to peak flux density (Y-factor) of $\leq$ 2, and further imposed S/N $>$ 10 to ensure reliable detection across most sub-images. We extracted the peak flux density and the uncertainty of each source in eight frequency bins, excluding the sub-band with high noise. We then performed least squares fitting using the \texttt{scipy} function \textit{curve\textunderscore fit} to the linear equation, $\log S_\nu = \alpha \times \log \nu + \mathrm{c}$. 
    
    We estimated spectral indices for 18 of the 49 extracted sources, 7 of which have $\alpha \geq -0.1$ within the uncertainty limits. The more reliable estimates of spectral indices are limited to the nine most compact sources with Y-factor $\leq$ 1.2. Of these nine sources, five sources have $\alpha \geq -0.1$, pointing toward thermal emission.   
    
    \subsection{Mid-infrared colors}
    \begin{figure}[h]
        \centering
        \includegraphics[height = 0.3\textwidth]{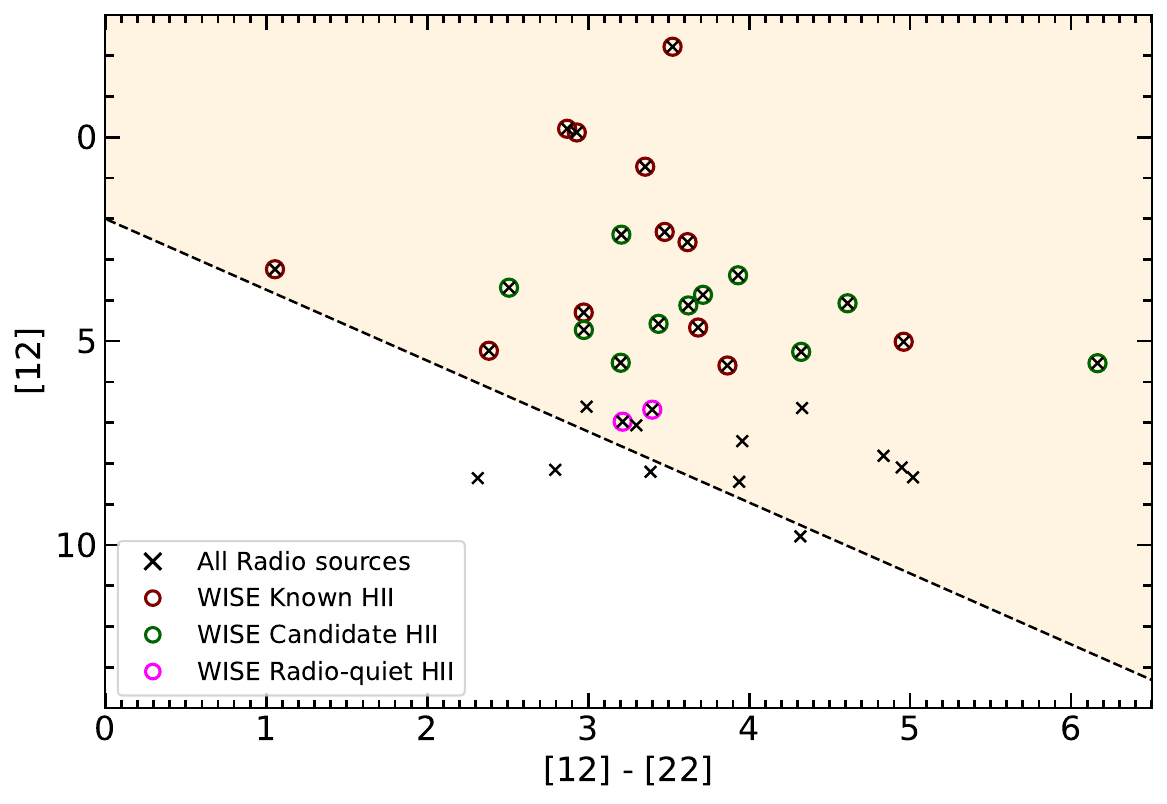}
        \caption{ MIR colors for the 37 sources (black crosses) associated with a WISE counterpart out of the total 49 radio sources found toward G358.69+0.03 in this work. The maroon, green, and magenta circles represent the known, candidate, and radio-quiet \ion{H}{ii} regions, respectively, from the WISE Galactic \ion{H}{ii} Regions Catalog V2.3 \citep{2014Anderson}. The region above the dashed black line represents the parameter space where all the \ion{H}{ii} regions and the majority of PNe are present, as identified by \cite{2019Sac}.} 
        \label{wise}
    \end{figure}
    \begin{figure*}[h]
        \centering
        \includegraphics[height =0.6\textwidth]{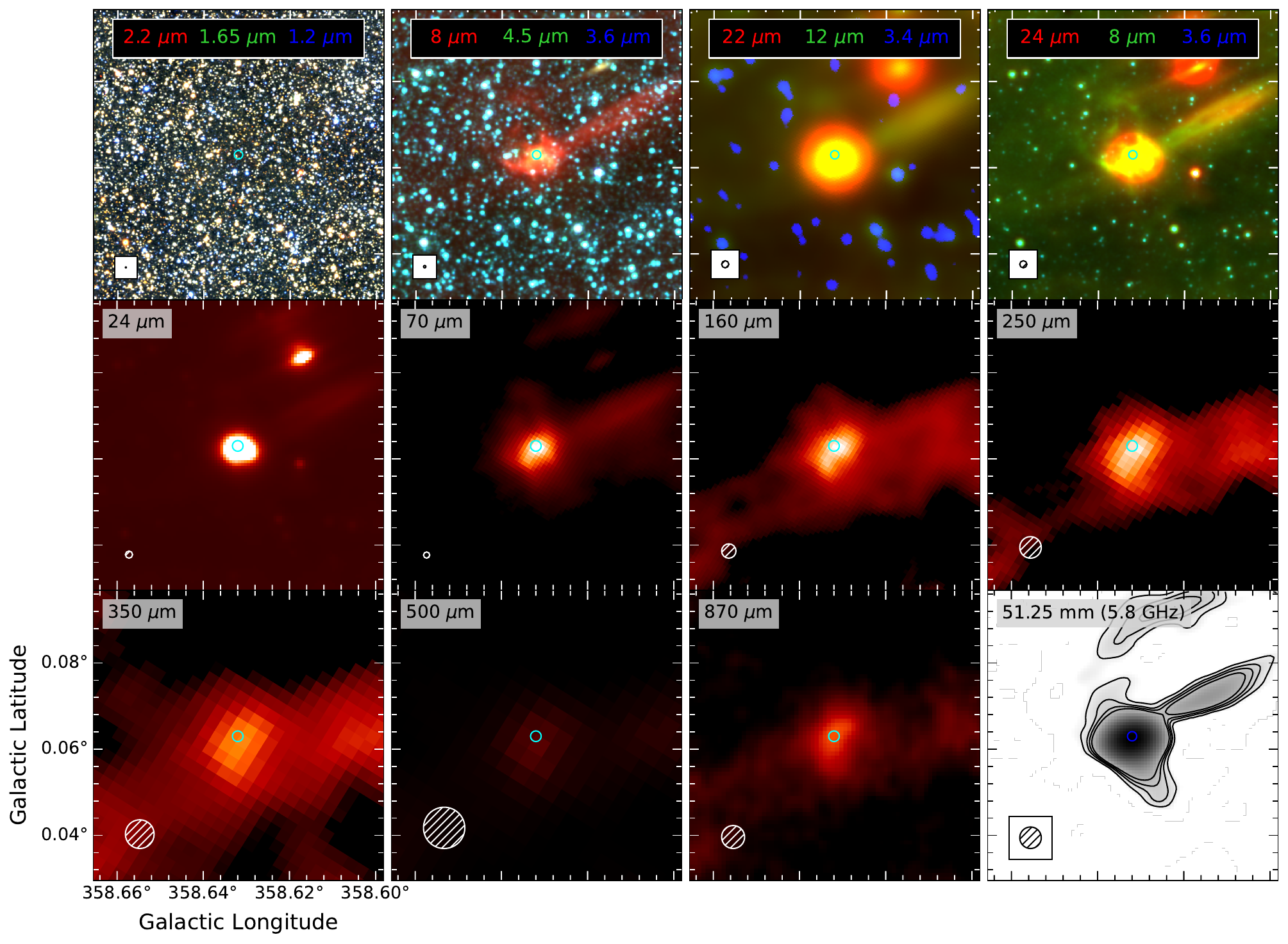}
        \caption{Multiwavelength view of the known \ion{H}{ii} region, G358.632+0.063. First row (from left to right): Color composite images from 2MASS \citep{2006Skrutskie}, GLIMPSE, WISE, and \textit{Spitzer}. The red, green, and blue colors are shown in each panel. Middle and bottom rows (from left to right and top to bottom): MIPSGAL 24 $\mu$m, Hi-GAL 70 $\mu$m, Hi-GAL 160 $\mu$m, Hi-GAL 250 $\mu$m, Hi-GAL 350 $\mu$m, Hi-GAL 500 $\mu$m, ATLASGAL 870 $\mu$m, and 5.8 GHz GLOSTAR VLA-D images of the \ion{H}{ii} region. The blue circles in the center of each panel show the peak position of radio emission. The FWHM beams of 2MASS (0.8$''$), GLIMPSE (2$''$), WISE (6$''$ at 12 $\mu$m), \textit{Spitzer} (6$''$ at 24 $\mu$m), MIPSGAL (6$''$ at 24 $\mu$m), Hi-GAL (6$''$, 12$''$, 18$''$, 24$'',$ and 35$''$ at 70, 160, 250, 350, and 500 $\mu$m, respectively), and ATLASGAL (19$''$) are indicated by the black and white circles with hatched lines shown in the lower-left corner of each panel.}
        \label{region1}
    \end{figure*}
    To further aid to the identification of c\ion{H}{ii} regions, we examined the MIR colors of our extracted sources using the classification scheme followed by \citet{2019Sac}. They generated a MIR color-magnitude diagram of the radio sources in the pilot region of the GLOSTAR survey (28\textdegree \ < \textit{l} < 36\textdegree, |\textit{b}| < 1\textdegree). They classified 231 continuum sources out of the total 1575 discrete radio sources as \ion{H}{ii} region candidates. They examined the MIR colors of the cataloged \ion{H}{ii} regions and planetary nebulae (PNe) from their study and from literature \citep{2013Purcell} and found a good agreement between the datasets. Dust embedded objects, such as \ion{H}{ii} regions and PNe, exhibit redder MIR colors and occupy a distinct region of the color-magnitude diagram compared to contaminants such as radio stars and extragalactic background sources. Based on that, they proposed a criterion to identify new \ion{H}{ii} region candidates and defined an area of parameter space using their color-magnitude plot where all of the \ion{H}{ii} regions and a vast majority of the PNe are located, which is given by
    \begin{equation} \label{eq} 
        [12] \mathrm{Mag}. < 1.74 \times ([12]-[22]) +2.
    \end{equation}
    
    We identified WISE counterparts using the WISE All-Sky Source Catalog \citep{2020Wise} and found cross-matches for 41 out of the total 49 radio sources within a 10$''$ search radius. Among these sources, 37 sources have reliable fluxes in all four WISE bands (3.4, 4.6, 12, and 22~$\mu$m), and their distribution in the color-magnitude diagram is shown in Fig.~\ref{wise}. Of these 37 sources, 33 lie within the parameter space proposed to be occupied by all the \ion{H}{ii} regions and a majority of PNe. Among these 33 sources, 12 correspond to known \ion{H}{ii} regions and 11 are listed as c\ion{H}{ii} regions in the WISE Galactic \ion{H}{ii} Regions Catalog V2.3 \citep[hereafter  the ``WISE \ion{H}{ii} catalog'';][]{2014Anderson}. Two additional sources are listed in the catalog as radio-quiet \ion{H}{ii} regions. The remaining 12 sources do not have a counterpart in the WISE \ion{H}{ii} catalog. We discuss the nature of these sources based on their detailed infrared and radio characteristics in Sect.~\ref{sec:class}.
    
    \subsection{Classification of \ion{H}{ii} regions}\label{sec:class}
    Based on our analysis, we cataloged radio sources as \ion{H}{ii} regions using multiple criteria such as the multiwavelength associations (MIR, FIR, and submillimeter), spectral index estimates, and the MIR color--magnitude diagram. We also looked for known, candidate and radio-quiet \ion{H}{ii} regions toward G358.69+0.03 using the WISE \ion{H}{ii} catalog. Among our 49 radio sources, we found 16 known and 11 c\ion{H}{ii} region cross-matches from the WISE \ion{H}{ii} catalog, which were classified as \ion{H}{ii} regions in our work provided they had multiwavelength counterparts. Figure~\ref{region1} shows an example multiwavelength view for one such known \ion{H}{ii} region, namely G358.632+0.063. We also found three sources, namely G358.505+0.108, G358.747--0.154, and G358.905+0.029, that had cross-matches with radio-quiet \ion{H}{ii} regions from the WISE \ion{H}{ii} catalog, which we then classified as c\ion{H}{ii} regions. The cross-matches with the WISE \ion{H}{ii} catalog are shown in Fig.~\ref{anderson}. 
    
    The remaining 19 sources were analyzed individually; among them, we identified 5 as new c\ion{H}{ii}. Table~\ref{hiianalysis} provides the detailed classification scheme used for their identification. All five new c\ion{H}{ii} regions occupy the same parameter space in MIR color-magnitude diagram where all the \ion{H}{ii} regions are located. Three of these sources, namely G358.612+0.178, G358.789--0.020, and G358.822+0.000, have $\alpha \geq$ --0.1 and at least 24 and 70 $\mu$m emission associated with them. Another source, G358.559--0.061, displays cold dust signature and is present at the edge of a radio-quiet \ion{H}{ii} region. Source G358.758+0.203 has compact emission at 8, 24 and 70 $\mu$m, with a negative spectral index of $\alpha$ = --0.34$\pm$0.15. Although \ion{H}{ii} regions are known to emit thermal bremsstrahlung radiation (Sect.~\ref{sec:si}), a number of observational studies have revealed the existence of \ion{H}{ii} regions with a mixture of thermal and nonthermal radiation, with steep negative spectral indices, \citep[e.g.,][]{2016Veena,2016AJ....152..146N,2019Meng,2023yang}. Using their model of thermal particle acceleration, \cite{2019Padovani} show that shocked \ion{H}{ii} regions can exhibit nonthermal emission, which might be due to synchrotron radiation from locally accelerated electrons braked in a magnetic field. Our region of interest is present at highly shocked and turbulent location, at the intersection of the CMZ and the far dust lane. The shocked gas is likely to give rise to the nonthermal emission in the target region. We discuss the presence of shock signatures in Sects. \ref{mir} and \ref{sioshocks}. 

\begin{figure}[h]
        \centering
        \includegraphics[height = 0.33\textwidth]{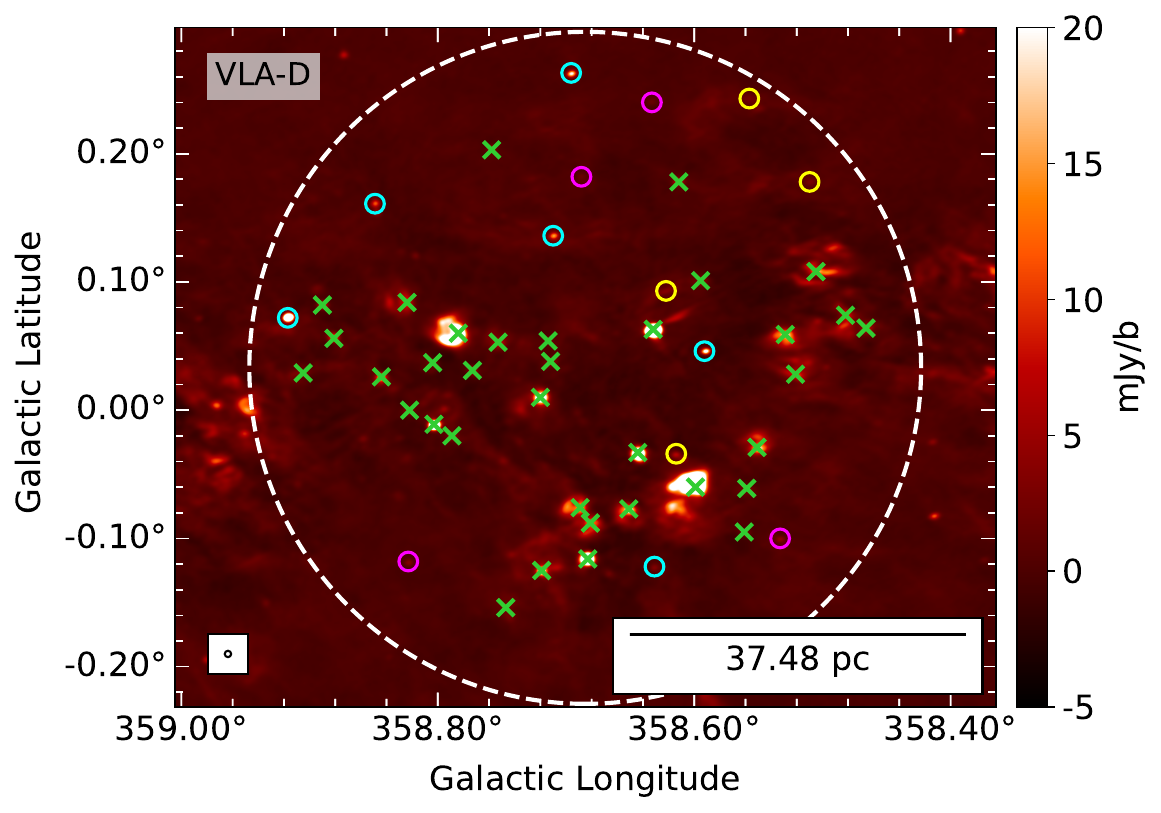}
        \caption{Classification of radio sources toward G358.69+0.03. The green crosses show the HII regions identified in this work. The cyan, magenta, and yellow circles mark the positions of cEGs, cPNe, and the unidentified radio (Rad) sources, respectively. The dashed white circle marks the target region. The background image shows the GLOSTAR VLA-D configuration image. The beam size is shown in the lower-left corner.}
        \label{img:source_class}
    \end{figure}

    The other 14 radio sources that are not classified as \ion{H}{ii} regions or c\ion{H}{ii} regions are listed in Table~\ref{propertiesnonhii}, along with their physical properties obtained from the \texttt{BLOBCAT} software. A few of these sources, namely G358.917+0.072, G358.696+0.263, G358.592+0.046, G358.849+0.161, G358.710+0.136, and G358.631--0.122, have bright and compact radio emission but do not have counterparts at wavelengths from the MIR to the submillimeter. They are also seen as point sources at NIR wavelengths and are likely to be extragalactic candidates \citep[cEGs;][]{2008ukidds, 2012cornish}. The remaining eight sources have no emission at the submillimeter wavelengths. In the FIR regime, they have no emission from 160 to 500~$\mu$m, and a very weak emission at 70~$\mu$m. Of these, four, namely G358.823--0.118, G358.533--0.100, G358.633+0.240, and G358.688+0.182 are seen as isolated point sources showing red emission in the NIR and MIR three color composite image. These are classified as candidate planetary nebulae \citep[cPNe;][]{2012A&A...537A...1Anderson, 2023Sergio}. The nature of the other four sources, two of which have extended radio emission, remains unclear and are hence classified as unidentified radio sources (Rad). Figure~\ref{img:source_class} illustrates the various types of radio sources toward G358.69+0.03, described above.
    
    \section{Results}\label{results}
    
    \subsection{Properties of candidate \ion{H}{ii} regions}
     
    Our analysis has identified a total of 27 \ion{H}{ii} regions and 8 c\ion{H}{ii} regions toward G358.69+0.03. Of the known 16 \ion{H}{ii} regions from the WISE \ion{H}{ii} catalog, 14 \ion{H}{ii} regions have line of sight velocities (v$_\mathrm{lsr}$) in the range --193.3 and --212.6~km/s making them likely members of Sgr\,E complex \citep[v$_\mathrm{lsr}$ $\sim$ --200 km/s;][]{1996Cram}, accounting for about 88\% sources. For these sources, we adopted a GC distance of 8.2 kpc \citep{2019Abuter} to estimate their physical properties.  For the remaining sources, although velocity information is unavailable to confirm their association with the G358.69+0.03 region, their 2D spatial association with confirmed \ion{H}{ii} region members supports adopting the same distance. Therefore, following the approach of \cite{2020Anderson}, we assumed a distance of 8.2 kpc for the remaining sources as well. Approximately 12\% of the \ion{H}{ii} regions ($\thicksim$ 4 sources) may lie at different distances along the lines of sight. To better constrain the distances and verify their association with G358.69+0.03, we plan to carry out much deeper follow-up observations of recombination lines toward these sources. We provide the details involved in estimating the physical properties in Appendix \ref{hii_properties}.

   \begin{figure}[h]
        \centering
        \includegraphics[height=0.33\textwidth]{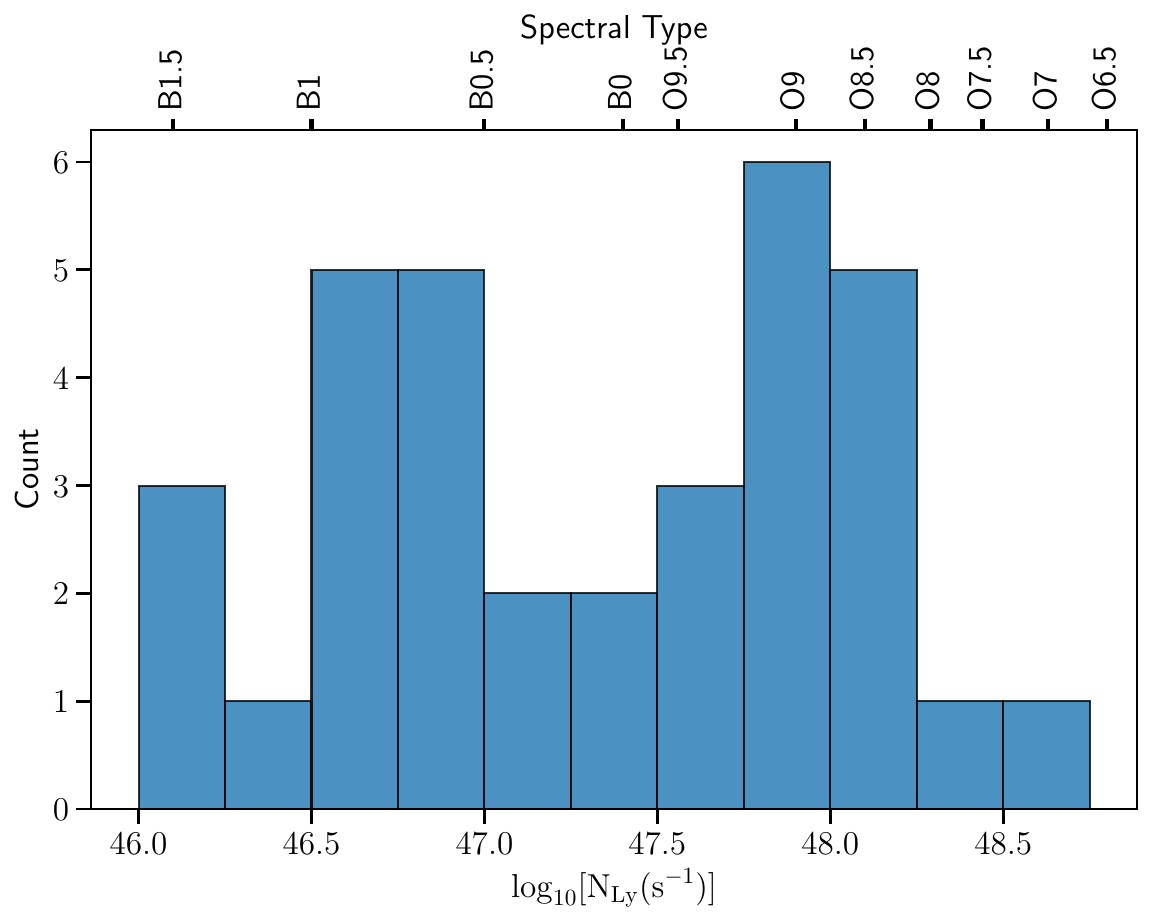}
        \caption{Spectral types and Lyman continuum photon rates of the \ion{H}{ii} regions derived from the VLA-D radio continuum data.}
        \label{spectraltype}
    \end{figure}

    The spectral types  of the central ionizing stars range from very massive O6.5 type stars to B1.5 type stars. The resultant distribution of the spectral types is shown in Fig.~\ref{spectraltype}. Table~\ref{propertieshii} provides all the detailed physical properties of the \ion{H}{ii} regions. The effective radii (0.35-3.87 pc) and the dynamical ages (0.07-0.94 Myr) of the identified \ion{H}{ii} and c\ion{H}{ii} regions are characteristic of compact \ion{H}{ii} region stage. Additionally, we observed that regions with the largest effective radii tend to have lower electron number densities and emission measure (EM) values, along with the highest dynamical ages, indicating a more evolved stage of \ion{H}{ii} region development. Notably, the oldest and largest \ion{H}{ii} regions are predominantly located on the western and southwestern periphery of the target region.  

    \begin{figure*}[!htbp]
        \centering
        \includegraphics[height=0.35\textwidth]{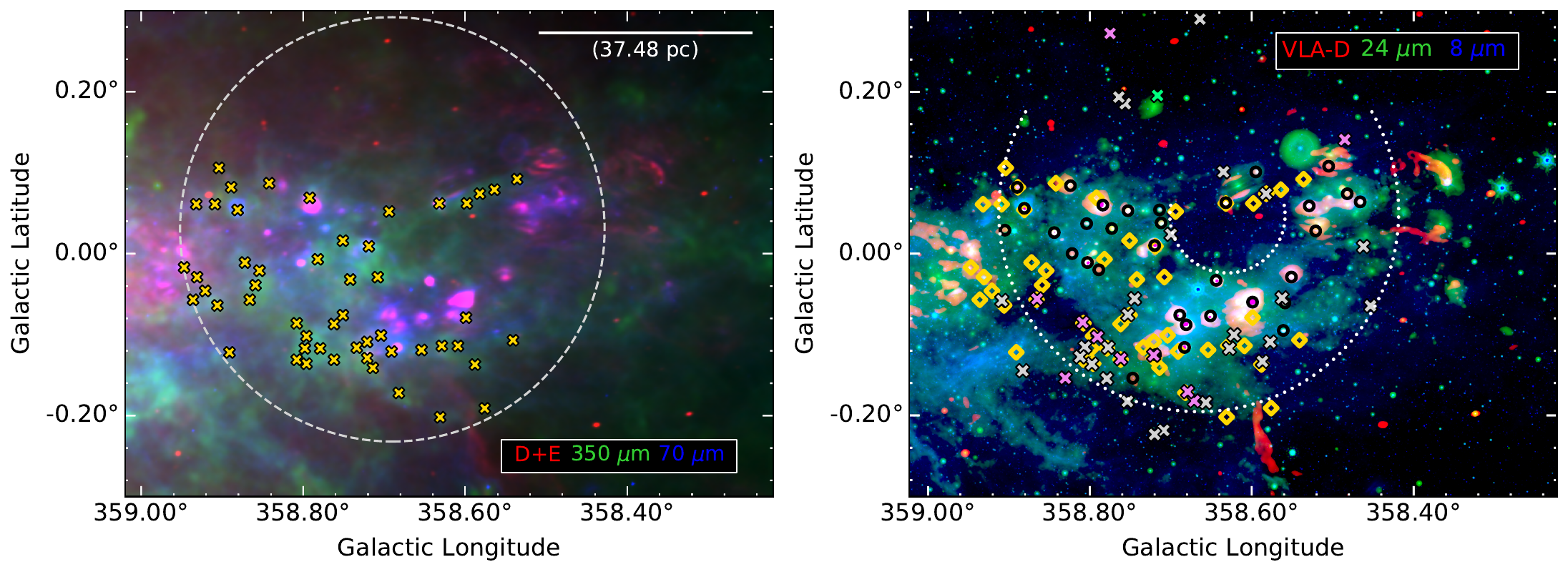}
        \caption{\textit{Left:} Distribution of the ATLASGAL dust continuum sources (gold markers) overlaid on top of the color composite image of G358.69+0.03. The VLA-D+Effelsberg image is shown in red and the Hi-GAL 350 and 70 $\mu$m emission maps in green and blue, respectively. The dashed white circle shows G358.69+0.03. \textit{Right:} Color composite image showing the VLA-D image in red and \textit{Spitzer} bands of 24 and 8 $\mu$m in green and blue, respectively. The dashed white partial circles mark the arc-like or U-shaped morphology of the target region. The black circles represent the \ion{H}{ii} regions identified in this work, and the gold diamonds mark the position of ATLASGAL cold dust clumps toward G358.69+0.03. The gray, violet, and green crosses mark the positions of pre-stellar clumps, protostellar clumps, and gravitationally unbound starless Hi-GAL clumps, respectively, at a distance of 8--9 kpc toward G358.69+0.03.}
        \label{atlasgal}
    \end{figure*}
    
    \cite{2020Anderson} reported that the \ion{H}{ii} regions population of the Sgr\,E complex exceeds 60, with an average Lyman continuum luminosity of log$_\mathrm{10}$($N_\mathrm{Ly}$) $\simeq$ 46.97. For the 35 \ion{H}{ii} regions in the G358.69+0.03 complex, which had cross-matches with 14 known and 10 c\ion{H}{ii} regions listed in \cite{2020Anderson}, we estimated a similar value of log$_\mathrm{10}$($N_\mathrm{Ly}$) $\simeq$ 47.37. Comparing with Sgr\,B2 molecular cloud, which hosts 54 \ion{H}{ii} regions source and is located on the positive longitude side of the GC at a projected distance of $\thicksim$100 pc, we found a comparable average log$_\mathrm{10}$($N_\mathrm{Ly}$) of 47.42, as reported by \cite{2022Meng}. Their study also identified central ionizing stars with spectral types of B0.5 to O6. 
    

       
    \subsection{Distribution of cold dust clumps}
       
    To investigate the cold dust emission toward G358.69+0.03, we used the ATLASGAL CSC by \citet{2014urquhart}. At the submillimeter wavelength of 870 $\mu$m, the thermal emission from dust is considered to be optically thin and contribution from free-free emission should be negligible. This makes ATLASGAL an excellent tool for studying the cold dust clumps that harbor star formation and estimating the column density and total clump mass. The CSC provides a database of $\thicksim$ 10163 cold and massive dense clumps toward the inner Galaxy, which also includes samples of the earliest embedded stages of high-mass star formation activity \citep{2014urquhart}. We extracted all the sources from the CSC and identified approximately 50 cold dust clumps in our target region. The left panel of Fig.~\ref{atlasgal} presents the distribution of these dense clumps relative to the radio emission and the FIR maps from the \textit{Herschel} Hi-GAL survey \citep{2016Molinari}. Although we cannot confirm distances for the ATLASGAL clumps directly, in Sect.\,\ref{sioshocks} we present velocity-resolved SiO observations that demonstrate these clumps are kinematically associated with the G358.69+0.03 region. It is evident that the majority of cold dust clumps are concentrated toward the south and southeast of the G358.69+0.03 complex. Only five c\ion{H}{ii} regions are associated with an ATLASGAL counterpart within a search radius of 18$''$. 

    \begin{figure}[!tbp]
        \centering
        \includegraphics[height=0.3\textwidth]{Figures/NewML_higalclumps.pdf}
        \caption{Bolometric luminosity versus mass of Hi-GAL compact sources at distances between 8 and 9 kpc toward G358.69+0.03. The clumps are color-coded by temperature. The markers with thick lines are sources with high reliability parameters taken from \citet{2021Elia}. The shapes of the markers (as shown in the legend) indicate different classification categories. The low-mass regime is taken from \citet{1996Saraceno}.}
        \label{bolometricml}
    \end{figure} 

    We also investigated emission from Hi-GAL \citep{2016Molinari}, which traces the earliest stages of star formation by tracing emission from cold dust (T < 20 K) in the Milky Way \citep{2017elia}. To study the distribution of Hi-GAL cold dust clumps, we used the Hi-GAL CSC by \citet{2021Elia}. The heliocentric radial velocities and distances to these Hi-GAL clumps were estimated by \citet{2021mege}, using \textit{starbird}, a \texttt{Python} program utilizing the VIALACTEA Knowledge Base \citep[VLKB;][]{2021mege}. We extracted all sources from the Hi-GAL CSC \citep{2021Elia} toward the direction of G358.69+0.03 that had a distance between 8 and 9 kpc, and we obtained 25 Hi-GAL clumps. The distribution of these clumps is shown in the right panel of Fig.~\ref{atlasgal}. We notice a higher concentration of these cold clumps toward the south and southeast directions of G358.69+0.03, similar to that of ATLASGAL clumps. We also investigated the star formation activity in these Hi-GAL clumps using the evolutionary classification discussed in \citet{2021Elia}. The clumps were classified into the following categories: ``protostellar'' clumps, which contain active star formation activity, and quiescent ``starless'' clumps, based on the presence and absence of 70 $\mu$m emission. The starless clumps were further subclassified as gravitationally bound ``pre-stellar'' and unbound clumps using Larson's third law \citep{2021Elia}. The gravitationally bound pre-stellar clumps have masses, $M(r)$ > 460 \(M_\odot\) $(r/\mathrm{pc})^{1.9}$ at radii $r$. Figure~\ref{bolometricml} shows the bolometric luminosity versus mass of Hi--GAL clumps toward our target region. The temperature of these clumps ranges from $\thicksim$ 9-29 K. A majority of these clumps are pre-stellar and a very few are protostellar clumps harboring active high-mass star formation.
    
   The 350~$\mu$m cold dust emission is extended and patchy across the region. Within the arc-like distribution of compact \ion{H}{ii} regions, the 350 µm emission is relatively faint or absent, while the brightest clumps are concentrated outside the arc, particularly toward the south and southeast (see the left panel of Fig.~\ref{atlasgal}). This morphology suggests that the immediate surroundings of the \ion{H}{ii} regions are relatively depleted in cold dust, consistent with heating and dispersal by massive star formation activity, although we refrain from claiming a strict quantitative anticorrelation.
    
    \subsection{Mid-infrared emission from dust, PAHs, and shocks} \label{mir}
    The MIR emission maps are valuable for distinguishing between various components in star-forming regions, such as stars, dust continuum emission, PAH emission, and shocked gas \citep[e.g.,][]{2010MNRAS.406..952S}. To investigate the properties of warm dust emission in G358.69+0.03, we utilized \textit{Spitzer} IRAC MIR maps at three wavelength bands: 3.6, 4.5, and 8.0 $\mu$m. The MIR color composite image (8.0 $\mu$m: red, 4.5 $\mu$m: green and 3.6 $\mu$m: blue) is presented in Fig.~\ref{mainrgb}. In addition to the detection of emission associated with compact \ion{H}{ii} regions, we observe large-scale diffuse emission at 8~$\mu$m spanning $\thicksim$ 71.5 pc. This large-scale emission exhibits a broken shell-like morphology that closely aligns with the distribution of c\ion{H}{ii} regions (see Fig.~\ref{target}), suggesting a potential association. The 8~$\mu$m emission may also include contributions from PAH molecules in the photodissociation regions (\citealt{{1984leger},{2009Churchwelll}}). The observed MIR shell-like structure points to an origin related to stellar feedback, such as stellar wind bubbles, evolved \ion{H}{ii} regions where the ionized gas has expanded  and is interacting with the ambient interstellar medium, or supernova remnants \citep[e.g.,][]{{2006AJ....131.1479R},{2008ApJ...681.1341W},{2010deharveng},{2011AJ....142...47P}}. We explore this in detail in Sect. \ref{discussion}. 
       
\begin{figure}[!tbp]
        \centering
        \includegraphics[height=0.3\textwidth]{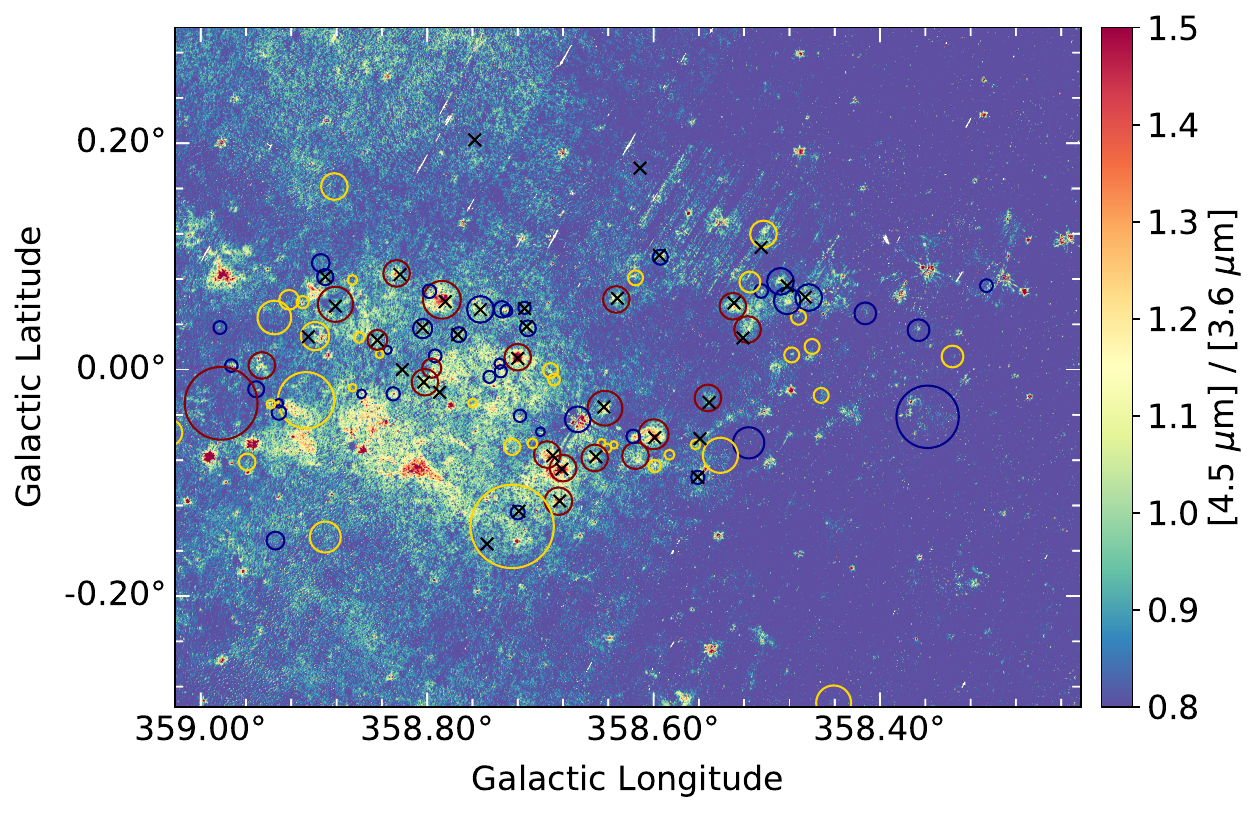}
        \caption{Flux ratio map of \textit{Spitzer} [4.5 $\mu$m]/[3.6 $\mu$m] for the target region. The black crosses show the \ion{H}{ii} regions identified in our work. The maroon, blue, and gold circles mark the known, candidate, and radio-quiet \ion{H}{ii} regions from the WISE \ion{H}{ii} catalog.}
        \label{ratio}
    \end{figure}

    Apart from the emission from heated dust, the 4.5~$\mu$m band also contains shock-excited lines of H$_2$ and CO \citep[][and references therein]{2013Liu}, making it a useful diagnostic for identifying shocks. To investigate this, we generated a pixel-by-pixel [4.5~$\mu$m]/[3.6~$\mu$m] flux ratio map for the region. The nearly identical point response functions of these bands ($\sim1.7''$) allowed us to forgo any point spread function matching correction. The resultant map is shown in Fig.~\ref{ratio}. A flux ratio $\geq$ 1.5 is typically associated with shocked regions, whereas stellar sources exhibit flux ratios $\ll$ 1.5 \citep{2010Takami,2018Veena}. From the map, we observe enhanced flux ratios ($\thicksim$2-3) primarily toward the eastern and southeastern portions of the arc, with several peaks aligning with c\ion{H}{ii} regions. This suggests significant shocked emission from these regions, likely due to stellar winds or the expansion of the \ion{H}{ii} regions. Apart from the high flux ratios near the peaks of c\ion{H}{ii} regions, we also observe regions with high flux ratios that lack associated radio emission. These could potentially be due to shocks from objects in early evolutionary phases, where the ionized gas has not yet produced strong radio emission, or alternatively, from shocks in molecular outflows or other dynamic processes in less evolved regions.       
    
    \subsection{Molecular gas as a tracer of kinematics and shocks }\label{sioshocks} 
    \begin{figure*}[h]
        \centering
        \hspace*{-1cm}\includegraphics[height=0.33\textwidth]{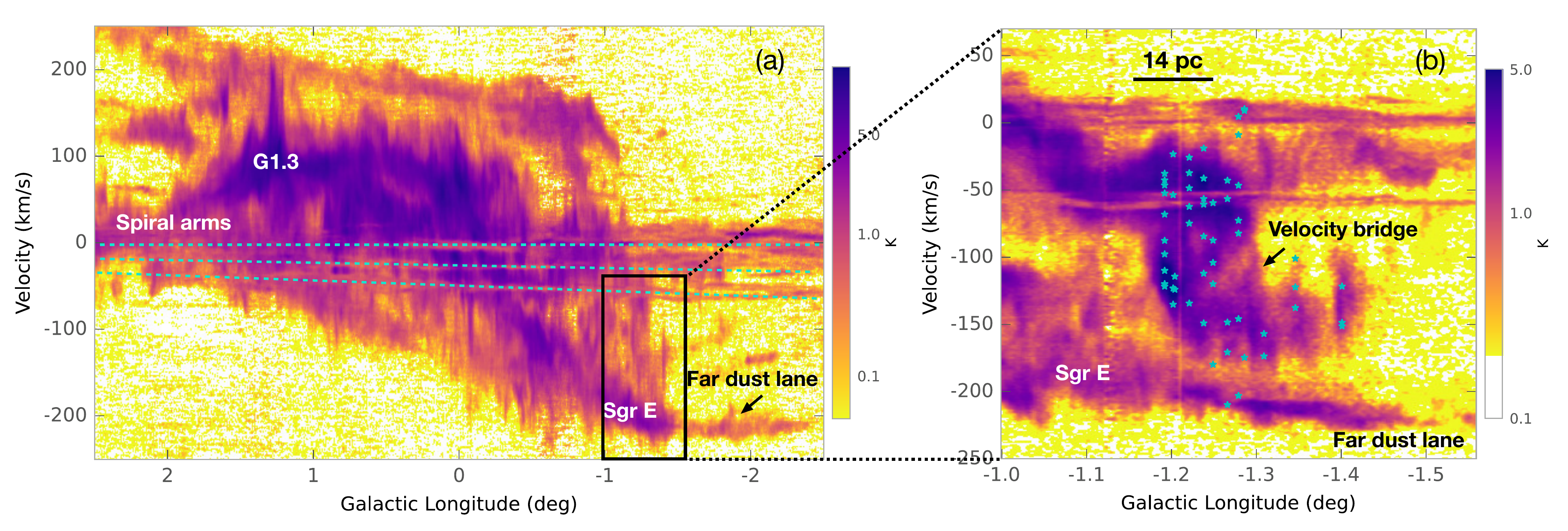}
        \hspace*{-1cm}
        \caption{\textit{Left:} Kinematics of the GC region revealed by the $^{12}$CO(3--2) longitude-velocity diagram. \textit{Right:} Zoomed-in view of the LV diagram of the G358.69+0.03 region (\textit{l}$\sim$ --1.3$^\circ$ or 358.7$^\circ$). Cyan stars correspond to the 57 components of the SiO emission detected toward the 15 ATLASGAL clumps listed in Table \ref{sio}. The scale bar corresponds to the assumed GC distance of 8.2~kpc.}
        \label{sio_pv}
        \end{figure*}
    To investigate the large-scale dynamics of the GC and its potential impact on high-mass star formation in G358.69+0.03, we analyzed $^{12}$CO($J$=3–2) data from the CHIMPS2 survey. Figure~\ref{sio_pv} shows longitude–velocity (LV) diagrams. In the large-scale LV diagram of the CMZ (Fig.~\ref{sio_pv}a), several well-known features are evident, including the spiral arms at positive velocities, the G1.3 cloud \citep{2022A&A...668A.183B} , and the far dust lane at negative velocities. Sgr\,E is located at the interface of the far dust lane and the CMZ orbit \citep{2020Anderson,2022ApJ...939...58W}, coinciding with a prominent stream of high-velocity emission. The zoom-in LV diagram of the G358.69+0.03 region (Fig.~\ref{sio_pv}b) reveals a striking velocity bridge linking the far dust-lane gas ($v_{\mathrm{LSR}} \lesssim -190$ km\,s$^{-1}$) with the main CMZ component ($-180 \lesssim v_{\mathrm{LSR}} \lesssim 0$ km\,s$^{-1}$). This bridge spans more than 100 km\,s$^{-1}$ in velocity and is an indicator of interaction between Galactic bar driven dust lanes and the CMZ \citep{2019MNRAS.488.4663S}. The location of compact \ion{H}{ii} regions in this interface further suggests that the observed massive star formation activity could be related to this dynamical interaction.

\begin{figure}[!htbp]
        \centering
        \includegraphics[height=0.3\textwidth]{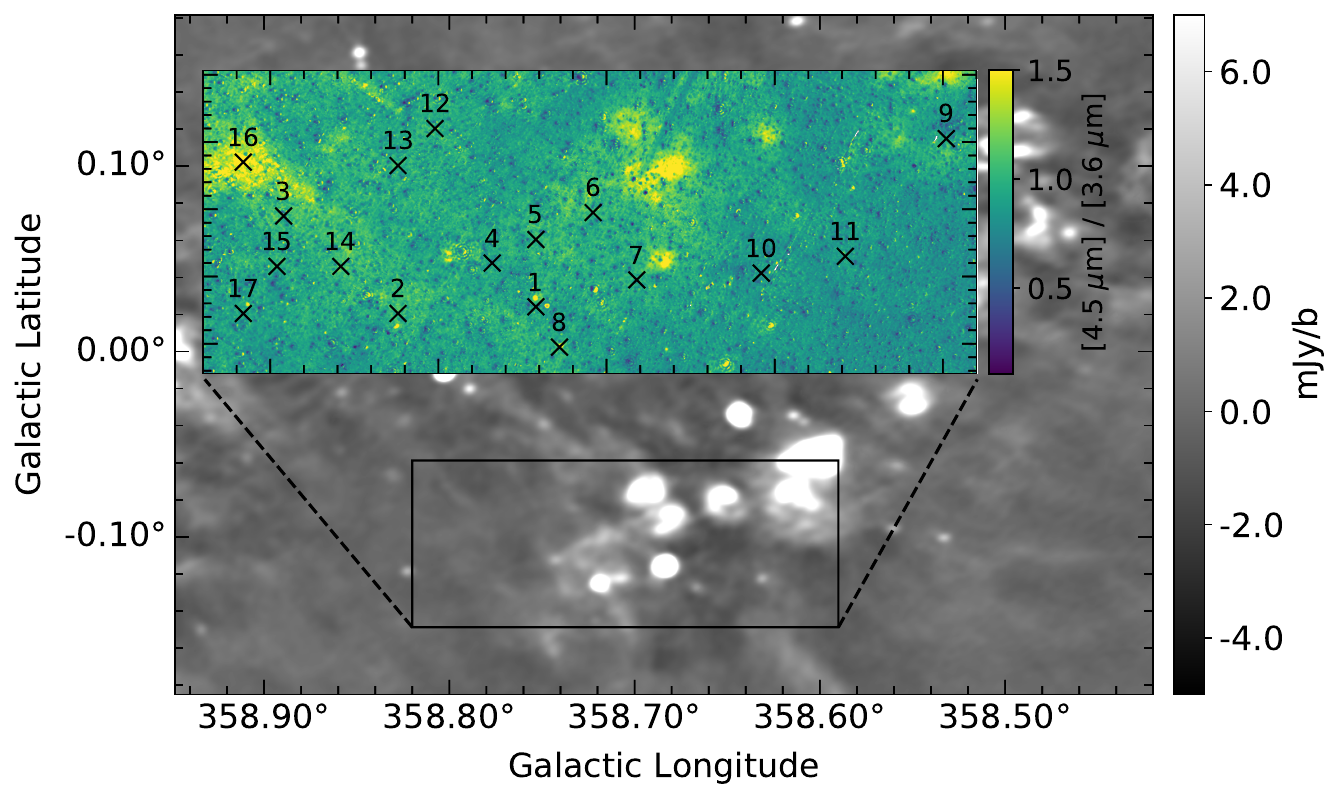}
        \caption{Sources observed with the APEX nFLASH230 receiver at a tuning frequency of 230.79 GHz (black crosses) overplotted on the inset flux ratio map of \textit{Spitzer} [4.5 $\mu$m]/[3.6 $\mu$m] for the southern part of the target region, G358.69+0.03. The numbers represent the corresponding IDs of the sources given in Table \ref{sio}. The background image shows the VLA-D continuum map. }
        \label{apexsources}
    \end{figure}

    To test whether the velocity bridge is indeed associated with shocked gas, we targeted dense clumps south of the arc-like distribution of \ion{H}{ii} regions (\textit{l},\ \textit{b}=358.70\textdegree,$-$0.11\textdegree; 15\arcmin$\times$5\arcmin). Within this region, 20 ATLASGAL clumps are cataloged, of which the 17 brightest were observed. We carried out SiO($J$=5–4) observations toward these 17 clumps (Fig.~\ref{apexsources}). SiO is a well-established tracer of shocks, being enhanced both in protostellar outflows and in cloud–cloud collisions \citep[e.g.,][]{1998Gueth,2010Jim,2023A&A...679A.123K}. SiO emission is detected in 15 out of the 17 clumps, with all detected spectra exhibiting more than one emission component (Fig.~\ref{spectrum_of_AGAL_sources}). The fitted line parameters are listed in Table~\ref{sio}. The local standard of rest velocities of the components range between –209.9 and +10.4 km\,s$^{-1}$, with line widths spanning 3.2–53.4 km\,s$^{-1}$ and an average of $\sim$16 km\,s$^{-1}$. Nine clumps exhibit very broad components ($>20$ km\,s$^{-1}$) characteristic of high-velocity shocked gas \citep[e.g.,][]{2019ApJ...878...29L}. In addition, several spectra display wing-like profiles. 
    
    While some of the SiO emission can arise from protostellar outflows, the prevalence of multiple, well-separated velocity components in $>65\%$ of the clumps indicates that large-scale shocks, such as those induced by cloud–cloud collisions, must play a significant role in this region. Although the recombination line velocities of 14 known \ion{H}{ii} regions from \citet{2020Anderson} cluster around --217 to --193 km\,s$^{-1}$, the molecular clumps in their vicinity exhibit a much broader spread, extending from --210 to +10 km\,s$^{-1}$. Notably, 13 of the 15 SiO-detected clumps exhibit at least one velocity component at $v_{\rm LSR} \leq -120$ km\,s$^{-1}$, well outside the range of foreground spiral arms \citep[e.g.,][]{2016ApJ...823...77R}, confirming their association with the CMZ. A large velocity spread in SiO is a natural outcome in the CMZ, where colliding inflows produce shocked molecular gas at a range of velocities \citep[e.g.,][]{2009ApJ...692...47M}. 

    \begin{figure*}[!tbp]
        \centering
        \hspace*{-0.5cm}
        \includegraphics[height=0.4\textwidth]{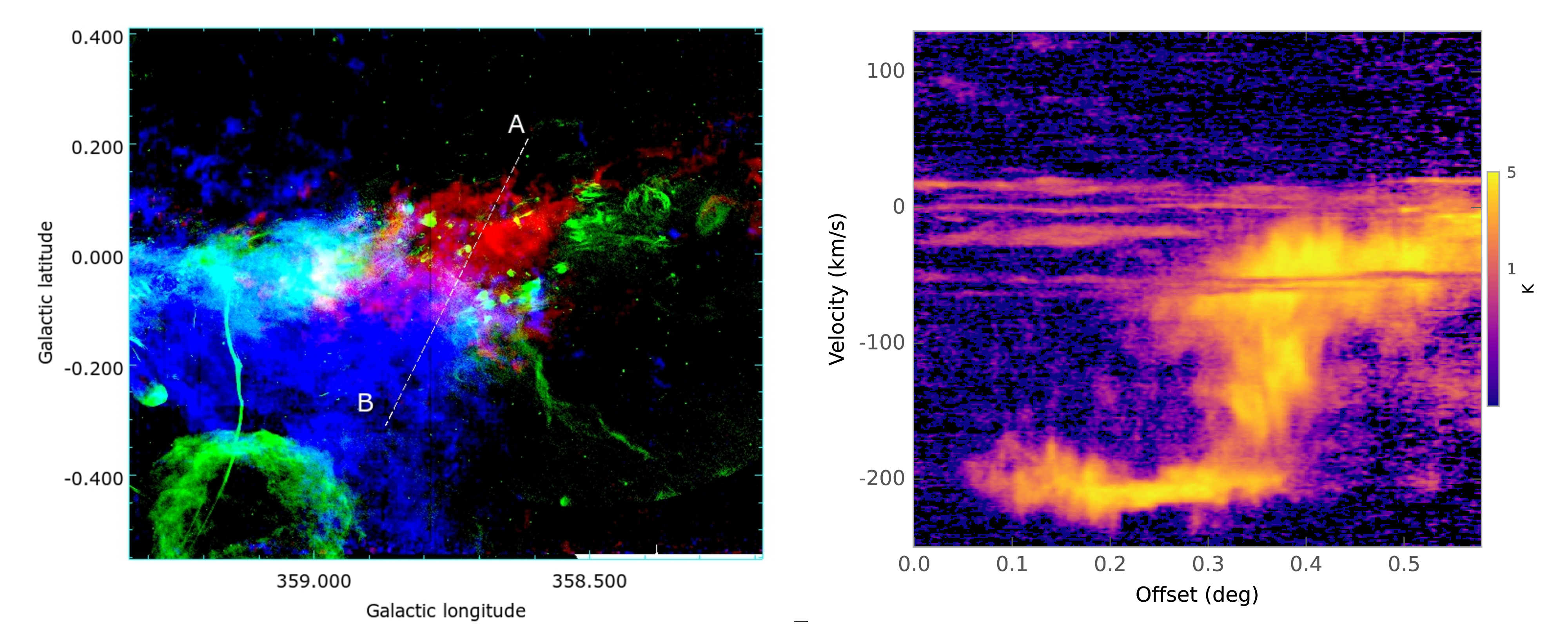}
        \caption{\textit{Left:} Three-color map of the Sgr\,E region. Red shows $^{12}$CO($J$=3–2) emission integrated over –240 to –190 km\,s$^{-1}$, tracing the far dust-lane component; blue shows –185 to –50 km\,s$^{-1}$, tracing the main CMZ stream and the emission bridge observed in Fig.~\ref{sio_pv}(b); and green corresponds to the MeerKAT 1~GHz image, tracing the \ion{H}{ii} regions. The dashed line (AB) indicates the cut used to construct the LV diagram. \textit{Right:} $^{12}$CO($J$=3–2) LV diagram along the AB line, revealing a velocity bridge linking the far dust-lane gas with the CMZ stream, consistent with a site of cloud–cloud interaction.}
        \label{RGBpv}
    \end{figure*}

    Under the assumption that the SiO(5 -- 4) emission is optically thin and the gas is under local thermodynamic equilibrium, we estimated the SiO column densities toward individual clumps using the expression \citep[e.g.,][]{2023A&A...679A.123K}
    \begin{equation}
        \begin{split}
            N^{\mathrm{thin}}_{\mathrm{tot}}=\bigg(\frac{8\pi\nu^3}{c^3A_{ul}}\bigg)\,\bigg(\frac{Q(T)}{g_u}\bigg)\, \frac{\mathrm{exp}\big(\frac{E_u}{k_\mathrm{B}T_\mathrm{ex}}\big)}{\mathrm{exp}\big(\frac{h\nu}{k_\mathrm{B} T_\mathrm{ex}}\big)-1} \\
            \times \ \frac{1}{[J_\nu(T_\mathrm{ex})-J_\nu(T_\mathrm{bg})]}
            \int\frac{T_\mathrm{MB}}{f} dv\\
            \simeq 1.35 \times 10^{12} \ \int T_\mathrm{MB} \ dv
        \end{split}
    \end{equation}
    \noindent where $\nu$ = 217104.984~MHz is the rest frequency of the SiO $J=5-4$ line, $c$ is the speed of light, $A_{ul}$ = $5.296\times10^{-4} \ \mathrm{s}^{-1}$ is the Einstein coefficient for spontaneous emission, $Q(T)$ is the partition function, $g_u$ is the statistical weight of the upper state level, which is 11 for the (5 -- 4) line, and $k_\mathrm{B}$ is the Boltzmann constant. The energy of the upper level, $E_u$ is 31.26 K. All these values are taken from Cologne Database for Molecular Spectroscopy \citep[CDMS;][]{2005JMoSt.742..215M}. $\int T_\mathrm{MB}\,dv$ is the integrated intensity and $f$ is the beam filling factor, assumed to be unity. We considered an excitation temperature, $T_\mathrm{ex}$, of 25~K, which is equivalent to the average dust temperature in the region obtained from the \textit{Herschel} PPMAP method \citep{2017MNRAS.471.2730M}. The background emission temperature, $T_{\mathrm{bg}}$ is assumed to be 2.7 K corresponding to the cosmic microwave background radiation. The Rayleigh-Jeans temperature, $J_{\nu}(T)$ is given by $\frac{h\nu /k_{\mathrm{B}}}{\mathrm{exp}(h\nu / k_\mathrm{b}T)-1}$.  Using the above expression we find column densities ranging between $6.8\times10^{11}$~cm$^{-2}$ and $1.6\times10^{13}$~cm$^{-2}$. To estimate the abundance of SiO ($X(\textrm{SiO})$) with respect to $\textrm{H}_2$, that is, $N(\textrm{SiO})/N(\textrm{H}_2)$, we used the column density map of the region produced from the \textit{Herschel} data (70--500~$\mu$m) using the PPMAP method \citep{2017MNRAS.471.2730M}. The SiO abundances given in Table~\ref{SiOres} range from $5.3\times10^{-12}$ to $2.6\times10^{-10}$ with a median abundance of $3.8\times10^{-11}$. Obtained SiO column densities and abundances are typical of Galactic clouds with SiO detection \citep[e.g.,][]{{2016A&A...586A.149C},{2023A&A...679A.123K}}. 

    \begin{table}[!htbp]
        \centering  
        \caption{SiO column densities and relative abundances.}
        \begin{tabular}{c c c c}
            \hline
            Source&N(SiO)&N(H$_2$)&$X$(SiO) \\
            &cm$^{-2}$&cm$^{-2}$&\\
            \hline
            G358.721--0.129&$7.9\times10^{11}$&$1.5\times10^{23}$&$5.3\times10^{-12}$\\
            G358.762--0.131&$2.5\times10^{12}$&$6.7\times10^{22}$&$3.8\times10^{-11}$\\
            G358.796--0.102&$1.6\times10^{13}$&$6.2\times10^{22}$&$2.6\times10^{-10}$\\
            G358.734--0.116&$1.9\times10^{12}$&$5.1\times10^{22}$&$3.8\times10^{-11}$\\
            G358.721--0.109&$6.8\times10^{11}$&$4.8\times10^{22}$&$1.4\times10^{-11}$\\
            G358.691--0.121&$1.0\times10^{12}$&$3.1\times10^{22}$&$3.3\times10^{-11}$\\
            G358.714--0.141&$1.1\times10^{12}$&$1.2\times10^{23}$&$9.4\times10^{-12}$\\
            G358.599--0.079&$4.2\times10^{12}$&$2.8\times10^{22}$&$1.5\times10^{-10}$\\
            G358.654--0.119&$1.7\times10^{12}$&$3.3\times10^{22}$&$5.2\times10^{-11}$\\
            G358.751--0.076&$1.0\times10^{12}$&$4.2\times10^{22}$&$2.4\times10^{-11}$\\
            G358.762--0.087&$1.2\times10^{12}$&$4.7\times10^{22}$&$2.5\times10^{-11}$\\
            G358.779--0.117&$2.8\times10^{12}$&$6.8\times10^{22}$&$4.1\times10^{-11}$\\
            G358.798--0.117&$8.9\times10^{12}$&$5.8\times10^{22}$&$1.5\times10^{-10}$\\
            G358.808--0.086&$1.5\times10^{12}$&$1.5\times10^{23}$&$9.9\times10^{-12}$\\
            G358.808--0.131&$8.5\times10^{12}$&$5.8\times10^{22}$&$1.5\times10^{-10}$\\
            \hline
        \end{tabular}
        \label{SiOres}
    \end{table}

    \section{Discussion}\label{discussion}  
    From our analysis, it is clear that the G358.69+0.03 region is a particularly special part of the Sgr\,E complex. We identified 35 compact \ion{H}{ii} regions that follow a characteristic arc-like or U-shaped distribution. Multiwavelength data further reveal that diffuse MIR emission traces the same arc, while cold dust clumps and molecular gas are distributed around it, indicating ongoing star formation across multiple evolutionary stages. 
    
    The G358.69+0.03 region lies at the intersection of the far dust lane and the CMZ orbit, a location associated with some of the highest line-of-sight velocities in the Galaxy \citep{1996MNRAS.280.1110C}. This dynamical environment of the target region is significant because it channels gas inflows along the Galactic bar into the CMZ \citep{2020MNRAS.499.4455T,2021ApJ...922...79H}. Similar inflow–CMZ interactions are seen elsewhere in the GC. For example, ALMA observations of the G5 cloud at $(\ell, b)=(+5.4,-0.4)$ revealed a velocity bridge between components at $\sim$50 and $\sim$150 km\,s$^{-1}$, direct evidence of a dust-lane collision \citep{2023ApJ...959...93G}. The 200 pc high-velocity cloud studied by \citet{2024A&A...689A.121V}, interpreted as near-side dust-lane material overshooting after brushing the CMZ, shows widespread SiO emission and ongoing massive star formation. In contrast, the G1.3 cloud, another dust-lane accretion site, shows no associated star formation activity \citep{2022A&A...668A.183B}. These comparisons highlight that dust-lane termination points can play diverse roles in regulating star formation, depending on the local conditions of the inflow–orbit interaction.
    
    \begin{figure*}[!htbp]
        \centering
        \hspace*{-0.5cm}
        \includegraphics[height=0.35\textwidth]{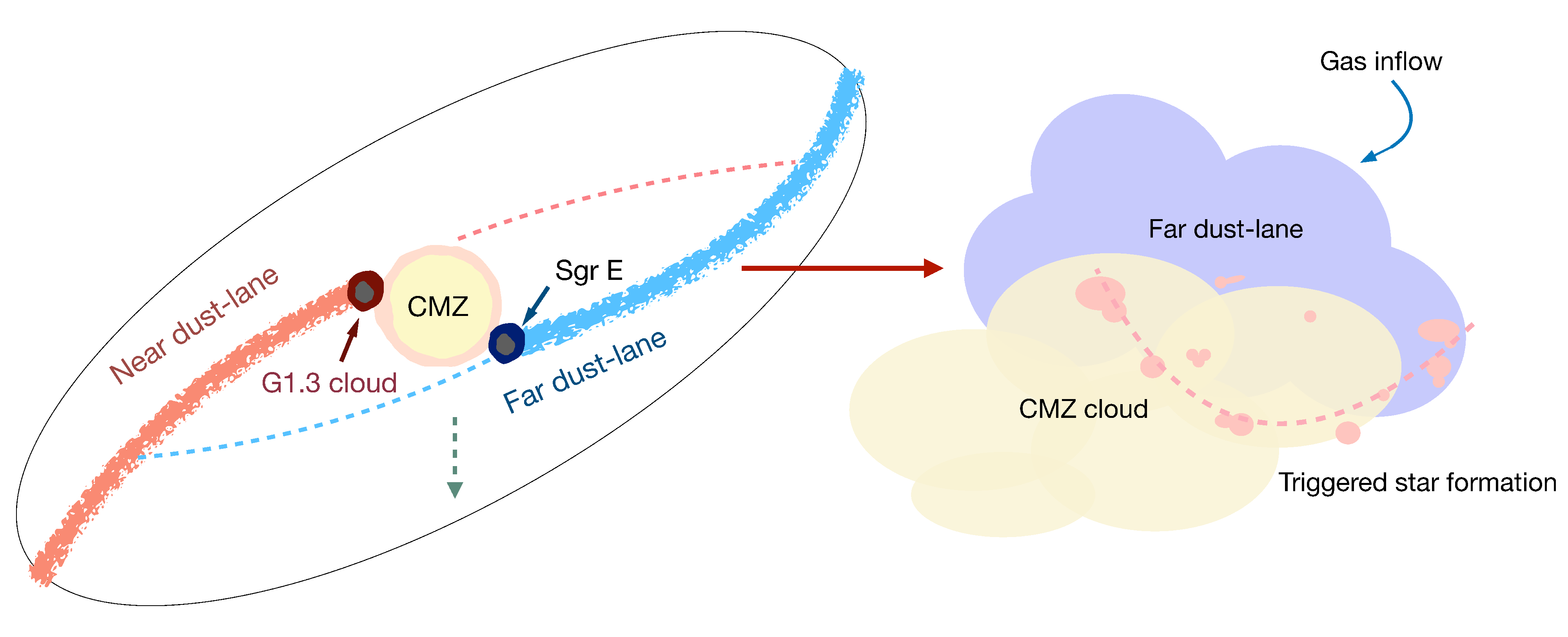}
        \caption{Schematic illustration of gas inflow into the CMZ. \textit{Left:} Gas streams along the near- and far-side dust lanes feed the CMZ. Sgr\, E is located at the intersection of the far dust lane and the CMZ orbit  \citep[adapted from][]{2022Henshaw}. \textit{Right:} Zoomed-in view of the collision, where the far dust lane impacts the CMZ gas, compressing material and producing an arc-like or U-shaped distribution of \ion{H}{ii} regions.}
        \label{schematic}
    \end{figure*}

    \subsection{Evidence of a dust-lane--CMZ interaction in G358.69+0.03}
    To investigate the interplay of large-scale gas inflow and dynamical interactions in G358.69+0.03, we constructed a color-composite image (see the left panel of Fig.~\ref{RGBpv}). The map reveals that most of the high-velocity CO emission is concentrated around the arc-like distribution of \ion{H}{ii} regions and the central cavity-like area enclosed by the arc. In contrast, the lower-velocity emission is predominantly located to the east and southeast of the \ion{H}{ii} arc, with only partial overlap with the far dust-lane component. Strikingly, the majority of the compact \ion{H}{ii} regions and ATLASGAL clumps (see Fig.~\ref{atlasgal}) are found precisely at this overlap zone, strongly suggesting that their formation is linked to the dynamical interaction between the two velocity components.

    To quantify this spatial correspondence, we constructed an additional LV diagram along the slanted line AB, which intersects the high-velocity far dust-lane emission, the lower-velocity CMZ gas, and their overlap region. The resulting diagram is shown in the right panel of Fig.~\ref{RGBpv}. In this LV diagram, we observe a characteristic J-shaped (or hook-like) feature that links the two velocity components. The coincidence of this kinematic bridge with the location of the \ion{H}{ii} regions provides strong evidence that the arc-like distribution of massive stars in G358.69+0.03 formed at the interface where the far dust-lane inflow terminates and collides with the CMZ stream.

    \subsection{Morphology of  the target region}
    Our kinematic analysis suggests that the morphology of G358.69+0.03 is a result of large-scale dynamical interactions. We investigated the two following scenarios in a bid to explain the morphology. 
    
    Feedback scenario: The structured star formation initially resembles an evolved feedback structure, such as an expanding \ion{H}{ii} region or supernova remnant, in which stellar feedback drives shell expansion and triggers secondary star formation at the rim \citep[e.g.,][]{2005Zavagno,2008Koo}. However, we could not find any evidence of a supernova remnant in this region. To further test the feedback scenario, we estimated the flux density of the large-scale ionized shell using the GLOSTAR VLA+Effelsberg map at 5.8 GHz, after subtracting diffuse background emission and compact sources. The total flux of 49 Jy corresponds to a Lyman continuum photon rate of $3.7\times10^{50}$ s$^{-1}$, consistent with ionization by an extremely massive O3-type star or a small OB cluster. If such a source existed, the expansion of an older \ion{H}{ii} region could in principle have compressed the surrounding molecular gas and triggered a new burst of star formation, consistent with sequential star-formation models. However, we find no direct observational evidence of a coherent expanding shell, and the recombination and supernova timescales ($\lesssim 10^5$ yr) are too short to account for the currently observed population of compact \ion{H}{ii} regions. 
    
    Could-cloud collisions scenario: On the other hand, the dynamical ages of the compact \ion{H}{ii} regions (0.06–1 Myr) are consistent with the $10^5-10^6$ yr timescales expected for massive star formation triggered by cloud–cloud collisions \citep[e.g.,][]{2011Torii}. The diffuse radio shell is therefore more plausibly explained as leakage of ionizing photons from the compact \ion{H}{ii} region complexes themselves, as observed in other Galactic star-forming regions \citep[e.g.,][]{2017ApJ...849..117L}. This scenario is summarized in Fig.~\ref{schematic}, which illustrates the inflow of gas along the near- and far-side dust lanes. Both lanes feed material into the CMZ, with G358.69+0.03 region located exactly at the critical intersection between the far dust lane and the CMZ orbit. At this termination point, as the shock wave moves outward from the point of collision, it can trigger widespread star formation in the compressed gas, leading to the formation of \ion{H}{ii} regions along the shock front. Numerical models of colliding clouds \citep{1992PASJ...44..203H,2014ApJ...792...63T,2021PASJ...73S...1F} also show that collisions between clouds of unequal density produce arc- or U-shaped compressed layers, with massive star formation occurring preferentially along the rim. This is consistent with the observed morphology of G358.69+0.03: a curved arc of compact \ion{H}{ii} regions, diffuse ionized gas filling the arc, and strong shock signatures traced by SiO.

    Cloud-cloud collision-driven star formation has been observed in many nearby interacting galaxies, including the Magellanic System and the Antennae Galaxies, as well as in almost all well-known high-mass star-forming regions in the Milky Way \citep{2021PASJ...73S...1F}. In the GC, \citet{2022ApJ...931..155E} investigated the effects of cloud-cloud collisions in the Sgr\,B region, one of the most active star-forming regions in the Galaxy. Their analysis revealed clear signatures of cloud-cloud collision, which they interpreted as the trigger for the vigorous star formation activity in the region.

    The cloud–cloud collision scenario provides a compelling explanation for both the morphology and the kinematics of G358.60+0.03. Furthermore, G358.69+0.03 marks the site where the far dust-lane inflow collides with the CMZ orbit, making it the key interaction zone within the broader Sgr\,E complex. This subregion thus provides a unique laboratory for studying how large-scale bar-driven gas inflows directly shape the morphology and trigger the formation of massive stars in the CMZ.

    \section{Conclusions}\label{conlusions}

    The G358.69+0.03 subregion of the Sgr\,E star-forming complex exhibits an intriguing arc-like morphology that is traced by diffuse 8~$\mu$m emission and a U-shaped distribution of compact radio and infrared sources. Using GLOSTAR 5.8 GHz VLA-D data, we identified 49 compact radio sources, 35  of which we classified as \ion{H}{ii} regions based on their spectral indices, associations with MIR or FIR sources, and corresponding cold dust emission. Among these, 27 were previously cataloged in the WISE V2.3 catalog \citep{2014Anderson}, while 3 additional radio counterparts are identified for previous ``radio-quiet'' WISE \ion{H}{ii} regions and 5 new c\ion{H}{ii} regions are reported. The derived spectral types span O6.5 to B1.5, with dynamical ages ranging from 0.06 to 1.05 Myr. We identified around 50 cold dust clumps in the region, primarily concentrated toward the south and southeast of the U-shaped \ion{H}{ii} region distribution, with evolutionary classifications ranging from pre-stellar to protostellar. The cold dust emission is relatively faint within the arc of compact \ion{H}{ii} regions, suggesting that massive star formation has already begun to alter the immediate environment, and active clump formation continues along the periphery. The MIR flux ratio maps reveal evidence of shocked emission, which is supported by widespread SiO detections in dense clumps. The SiO components span a broad velocity range and align with a continuous velocity bridge in CO, linking the high-velocity far dust-lane gas with the main CMZ stream.

    Given the location of G358.69+0.03 at the intersection of the far-side dust lane and the CMZ orbit, these results suggest that the observed star formation activity is shaped by large-scale dynamical interactions. The combination of high-velocity gas, shock tracers, and the arc-like distribution of \ion{H}{ii} regions is difficult to explain through stellar feedback alone. Instead, the evidence favors a scenario in which the collision between dust-lane inflow and the CMZ stream has compressed the gas and triggered massive star formation.

    \begin{acknowledgements}
    
    We would like to express our deepest gratitude to Prof. Dr. Karl Martin Menten for his guidance, and unwavering support and trust throughout this research. \\
    We thank the referees for the valuable comments and suggestions that improved the quality of the paper. This research was partially funded by the ERC Advanced Investigator Grant GLOSTAR (247078). A.C. acknowledges the support by the Deutsche Forschungsgemeinschaft (DFG, German Research Foundation) – Project-ID 500700252 – SFB 1601. This research has made use of the NASA/IPAC Infrared Science Archive, which is funded by the National Aeronautics and Space Administration and operated by the California Institute of Technology. This study utilized Astropy \citep{2022Astropy}, SAOImage DS9 \citep{2003ds9}, APLpy \citep{2012aplpy}, GIlDAS \citep{2005pety} and CASA \citep{2022Casa}. This document was prepared using the collaborative tool Overleaf available at: \url{https://www.overleaf.com/}. We thank S.-N.~X.~Medina for her support during the observation campaigns with Effelsberg. S.A.D. acknowledges the M2FINDERS project from the European Research Council (ERC) under the European Union's Horizon 2020 research and innovation programme (grant No 101018682). Y.G. was supported by the Ministry of Science and Technology of China under the National Key R\&D Program under grant No. 2023YFA1608200, the National Natural Science Foundation of China (NSFC) under grant No. 12427901, and the Strategic Priority Research Program of the Chinese Academy of Sciences under grant No. XDB0800301.  
    \end{acknowledgements}

    \bibliographystyle{aa.bst}
    \bibliography{bibliography.bib}
    
    \begin{appendix}
    \section{Properties of \ion{H}{ii} regions} \label{hii_properties}
    We estimated the physical properties of \ion{H}{ii} regions using the Str\"omgen sphere approximation, i.e., a fully ionized spherical region of uniform electron density and devoid of any interstellar dust. We also assumed the emission to be homogeneous and optically thin. The main source of radio flux from an \ion{H}{ii} region is the free-free thermal emission known as the bremsstrahlung radiation \citep{2016Condon}. The Lyman continuum luminosity ($N_{\mathrm{ly}}$), which represents the flux of ionizing photons emitted by the central massive star, is directly correlated with the stellar mass. The more massive a star, the higher is the production of ionizing photons and hence higher the values of $N_{\mathrm{ly}}$. Assuming that a single zero age main sequence star is responsible for the ionization of the \ion{H}{ii} region, we estimated the Lyman continuum photon rate under the assumption of optically thin emission, minimal leakage of photons into the interstellar medium and zero dust attenuation of the UV photons \citep[and references therein]{1973Panagia, 2017Armentrout, 2020Anderson}. $N_{\mathrm{ly}}$ was calculated using the expression \citep{1976AJ.....81..172Matsakis}
    \begin{equation} \label{lyman}
        \bigg(\frac{N_{\mathrm{ly}}}{\mathrm{s^{-1}}}\bigg) \ = \ 4.771 \times 10^{42} \times \bigg(\frac{F_{\nu}}{\mathrm{Jy}}\bigg) \  \bigg(\frac{T_{e}}{\mathrm{K}}\bigg)^{-0.45} \ \bigg(\frac{\nu}{\mathrm{GHz}}\bigg)^{0.1} \ \bigg(\frac{\mathrm{D}}{\mathrm{pc}}\bigg)^{2} \ ,
    \end{equation}
    where $F_\nu$ is the flux density of the \ion{H}{ii} regions as estimated from the GLOSTAR VLA-D radio continuum image using \texttt{BLOBCAT}. $T_e$ is the electron temperature which we assumed to be 6000 K for \ion{H}{ii} regions near the GC \citep{2011Balser, 2020Anderson}. The frequency $\nu$ =  5.8 GHz is the effective frequency of the GLOSTAR integrated intensity map. The distance (D) is taken to be 8.2 kpc \citep{2019Abuter}. The $N_{\mathrm{ly}}$ values are converted to main sequence spectral types using the data compiled in \citet{2017Armentrout}.  
    
    The effective radius of the c\ion{H}{ii} regions was estimated using 
    \begin{equation} \label{radiusHii}
        \bigg(\frac{R_{\mathrm{\ion{H}{ii}}}}{\mathrm{pc}}\bigg) \ = \ \frac{1}{57.32} \times \sqrt{\frac{n_{\mathrm{pix}} \times (\mathrm{pixel \ size})^2}{\pi}} \times \bigg(\frac{\mathrm{D}}{\mathrm{pc}}\bigg) \ ,
    \end{equation}
    where $n_{\mathrm{pix}}$ is the number of flooded pixels comprising the blob which was taken from the \texttt{BLOBCAT} output catalog. The size of the square pixels chosen to image the VLA-D configuration map was 2.5\arcsec. The effective radius ranges from 0.35-3.87 pc. The dynamical age of \ion{H}{ii} regions referring to the time taken by the \ion{H}{ii} region to evolve to its current size is estimated using \citep{1980DysonWilliams}
    \begin{equation} \label{tdyn}
        t_{\mathrm{dyn}} \ = \ \bigg(\frac{4R_{\mathrm{s}}}{7c_{\mathrm{s}}}\bigg) \ \Bigg[\bigg(\frac{R_{\mathrm{\ion{H}{ii}}}}{R_{\mathrm{s}}}\bigg)^{7/4} \ - \ 1\Bigg] \ ,
    \end{equation}
    where  $R_{\mathrm{\ion{H}{ii}}}$ is the effective radius of \ion{H}{ii} region. The velocity of sound in the ionized gas is taken to be $c_s$ = 10 km\,s$^{-1}$ \citep{2011Paron}, typical for \ion{H}{ii} regions. The radius of the Str\"omgen radius ($R_{\mathrm{s}}$) is estimated using the expression \citep{1939ApJ....89..526Stromgren}
    \begin{equation}\label{stromgenradius}
        R_{\mathrm{s}} \ = \ \bigg(\frac{3  N_{\mathrm{ly}}}{4  \pi  n_{0}^{2}  \alpha_{\mathrm{B}}}\bigg)^{1/3} \ ,
    \end{equation}
    where $n_0$ is the ambient particle number density of the neutral gas which is typically $\thicksim$ 10$^3$ cm$^{-3}$ \citep{2022Panja}. The hydrogen radiative recombination coefficient to all levels above the ground is given by $\alpha_B=2.6\times10^\mathrm{-13} \ (10^\mathrm{4} \ \mathrm{K}/T_e)^{0.7} \ \mathrm{cm}^3 \ \mathrm{s}^{-1}$ \citep{1997ApJ...489..284Kwan}. The dynamical age ranges from 0.07 to 0.94 Myr. The EM of an \ion{H}{ii} region is defined as $\mathrm{EM} = \int n_{\mathrm{e}}^2\, ds$, which is the fundamental quantity relating the electron density along the line of sight \citep{1967ApJ...147..471Mezger}. Assuming constant electron density, this expression simplifies to EM$=n^2_e\,L$, where $L$ is the path length through the region. Following \citet{2016Schmiedeke}, and adopting the formulation convenient for flux measurements within circular apertures, we calculated the EM and $n_\textrm{e}$ as  
    \begin{equation} \label{em}
        \bigg(\frac{\mathrm{EM}}{\mathrm{ pc \ cm^{-6}}}\bigg) \ = \ 3.217 \times 10^{7} \times \bigg(\frac{F_{\nu}}{\mathrm{Jy}}\bigg) \  \bigg(\frac{T_{e}}{\mathrm{K}}\bigg)^{0.35} \ \bigg(\frac{\nu}{\mathrm{GHZ}}\bigg)^{0.1} \ \bigg(\frac{\theta_{\mathrm{source}}}{\mathrm{arcsec}}\bigg)^{-2} ,
    \end{equation}
    \begin{equation} \label{ne}
    \begin{split}
        \bigg(\frac{n_{\mathrm{e}}}{\mathrm{cm^{-3}}}\bigg) \ = \ 2.576 \times 10^{6} \times \bigg(\frac{F_{\nu}}{\mathrm{Jy}}\bigg)^{0.5} \  \bigg(\frac{T_{e}}{\mathrm{K}}\bigg)^{0.175} \ \bigg(\frac{\nu}{\mathrm{GHZ}}\bigg)^{0.05} \\ \bigg(\frac{\theta_{\mathrm{source}}}{\mathrm{arcsec}}\bigg)^{-1.5} \ \bigg(\frac{\mathrm{D}}{\mathrm{pc}}\bigg)^{-0.5}
    \end{split}
    \end{equation}
    where $\theta_{\mathrm{source}}$ is the angular diameter of the \ion{H}{ii} region as estimated using the effective diameter at a distance (D) of 8.2 kpc. \citet{1967ApJ...147..471Mezger} showed that for alternative density distributions (cylindrical or Gaussian), the spherical case provides an upper limit to EM and a lower limit to $n_\textrm{e}$, but is generally consistent with these models. For the purposes of this study, we therefore retained the spherical model as a reasonable approximation, while noting that the true values may fall within the range implied by the different geometries. The EM values range from 2.64 to 35.2 $\times$ 10$^3$ pc cm$^{-6}$ and the $n_\mathrm{e}$ ranges from 29.48 to 121.20 cm$^{-3}$. 

    \subsection{Morphology of \ion{H}{ii} regions and their extended envelopes}
    \begin{figure*}[htbp]
    \centering
    \includegraphics[width=0.965\textwidth]{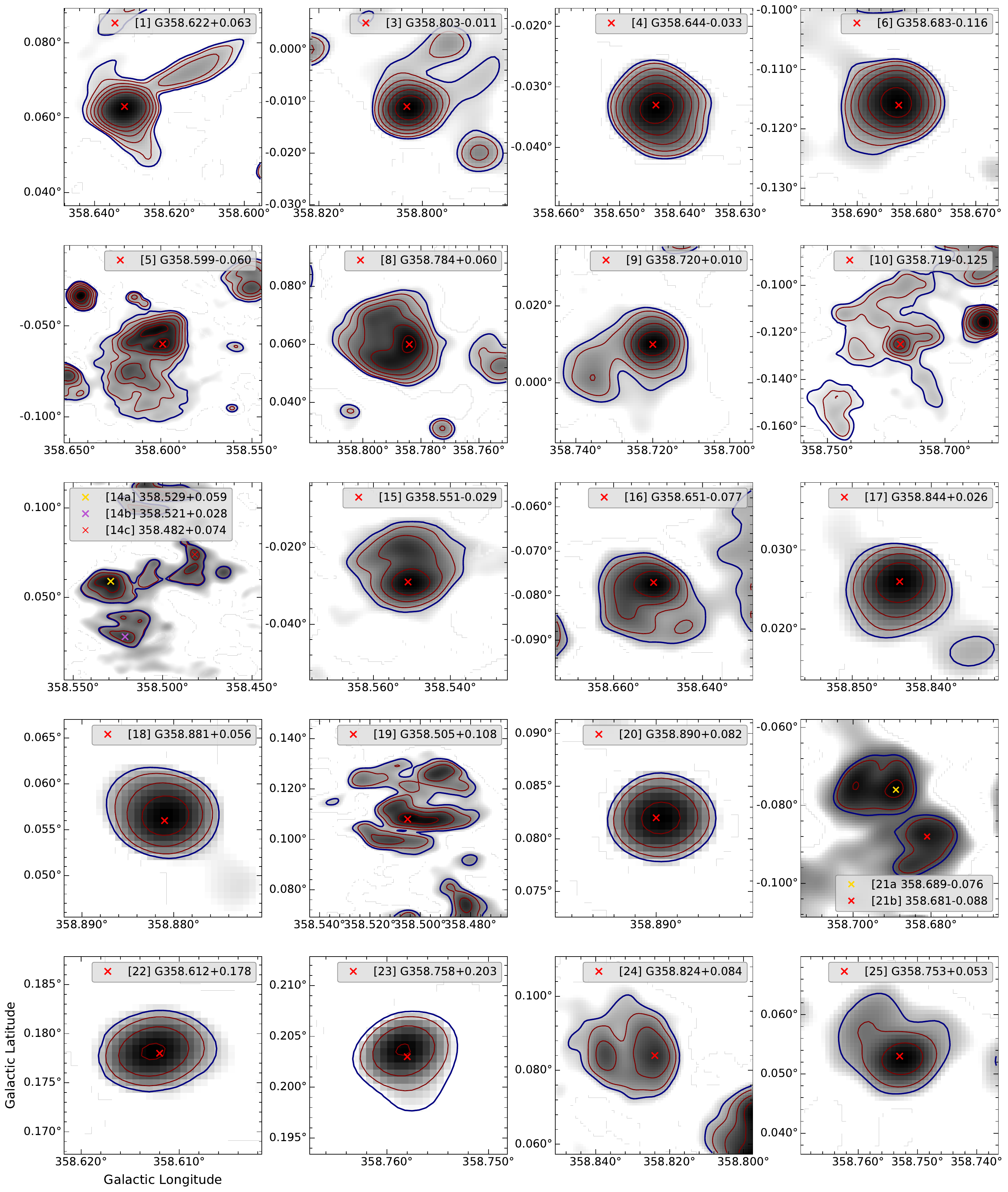}
    \caption{Morphologies of various \ion{H}{ii} regions. Each panel displays the source name and ID, along with the peak radio emission marked by cross symbols. The thick navy contours indicate the \texttt{BLOBCAT} regions used for the photometric measurements. They correspond to emission levels of 3\,$\sigma_{\mathrm{rms}}$ for all sources, except for sources 14 and 21, where the contours correspond to 4.6\,$\sigma_{\mathrm{rms}}$ and 7.6\,$\sigma_{\mathrm{rms}}$, respectively. The maroon contours represent emission levels of 5, 10, 20, 40, and 80\,$\sigma_{\mathrm{rms}}$, where the local $\sigma_{\mathrm{rms}}$ is obtained from the signal-to-noise ratio map created by \texttt{BLOBCAT}. Source 8 (G358.784+0.060) has a broken shell-like structure. Sources 5 (G358.599--0.060) and 15 (G358.551--0.029) show examples of the cometary morphology of \ion{H}{ii} regions. Source 19 (G358.505+0.108) is an example of multiple compact regions embedded in cometary-like diffuse gas.}
    \label{morphology}
    \end{figure*}
    
    \begin{figure*}[h]
    \ContinuedFloat
    \centering
    \includegraphics[width=0.965\textwidth]{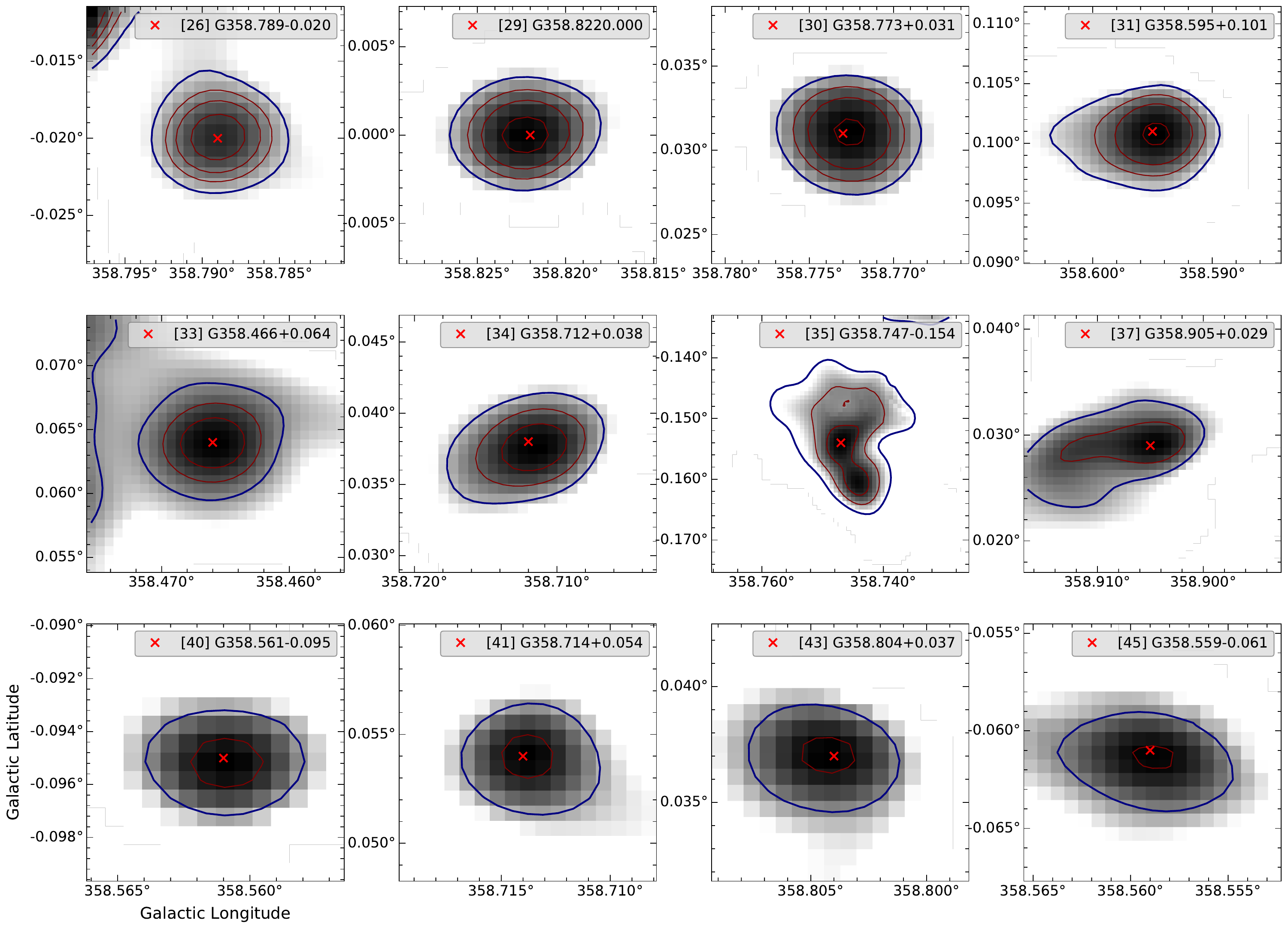}
    \caption[]{Continued.}
    \end{figure*}
    
    The \ion{H}{ii} regions identified in our study show different morphologies as shown in Fig.~\ref{morphology}. A few regions, such as G358.784+0.060, exhibit a broken shell-like structure, whereas some regions, such as G358.551--0.029 and G358.747--0.154, have a cometary morphology characterized by a cometary head and a low brightness tail. There are several irregular or multiple peaked \ion{H}{ii} region complexes, like G358.505+0.108, where multiple compact regions are clustered and embedded within a cometary-like low brightness diffuse gas. The \ion{H}{ii} region G358.632+0.063 is located near the geometric center of the shell and exhibits a distinct morphology. In addition to the central compact source, which has slightly extended emission in the north-south direction, it possesses a characteristic tail extending $\sim$3.7 pc in the southeast-northwest direction. This tail is also prominent in MIR and FIR emission (Fig.~\ref{region1}). The cold dust emission in this region appears filamentary, with its major axis aligned similarly to the ionized tail. The slight extension of the ionized envelope perpendicular to the cold dust filament and the ionized tail could be attributed to density gradients, where the ionized gas preferentially expands into lower-density regions, as observed in typical bipolar and blister-type \ion{H}{ii} regions \citep{{1979A&A....71...59T},{2015A&A...582A...1D}}. The ionized tail, on the other hand, could result from the motion of the ionizing source within the cloud, similar to the structures observed around runaway O and B stars. However, the observed tail does not exhibit the typical compressed, curved morphology of a classic bow shock \citep[e.g.,][]{2019arXiv190700122H} and might instead be shaped by photo-evaporative flows, interactions with large-scale gas motions, or the influence of magnetic fields, which could help channel ionized gas along preferred directions \citep[e.g.,][]{2013MNRAS.436..859M}. Further investigation of the kinematics of the ionized gas through RRL observations is necessary to distinguish between these scenarios. 
    \vspace{-0.5em}
    \section{Auxiliary data and figures} 
    \vspace{-1.5em}
    \begin{figure}[!htbp]
    \centering
    \includegraphics[height=0.3\textwidth]{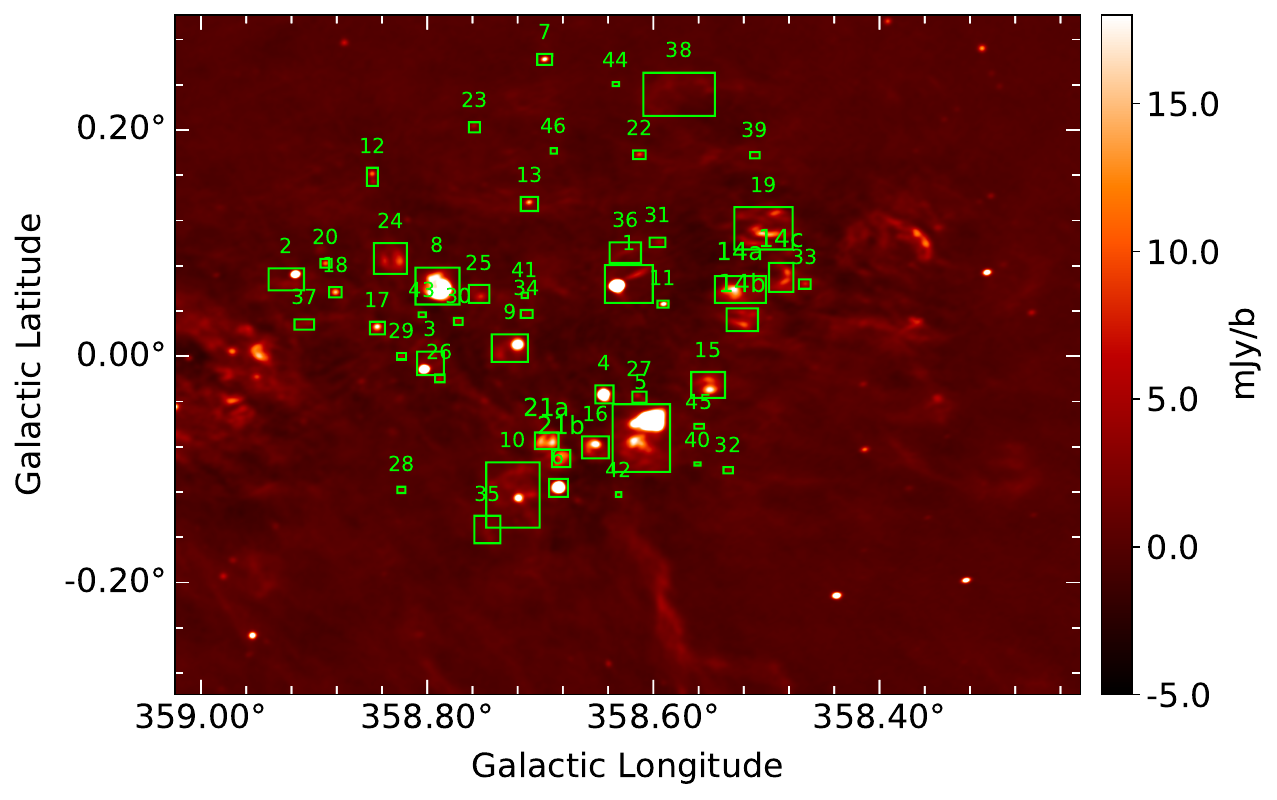}
    \caption{Sources extracted from the GLOSTAR VLA-D configuration image using \texttt{BLOBCAT}. They are shown as green boxes along with the assigned ID number given in Col. 1 of Tab.~\ref{count}.}
    \label{blobcat}
    \end{figure}

    \begin{figure}[!tbp]
        \centering
        \includegraphics[height = 0.33\textwidth]{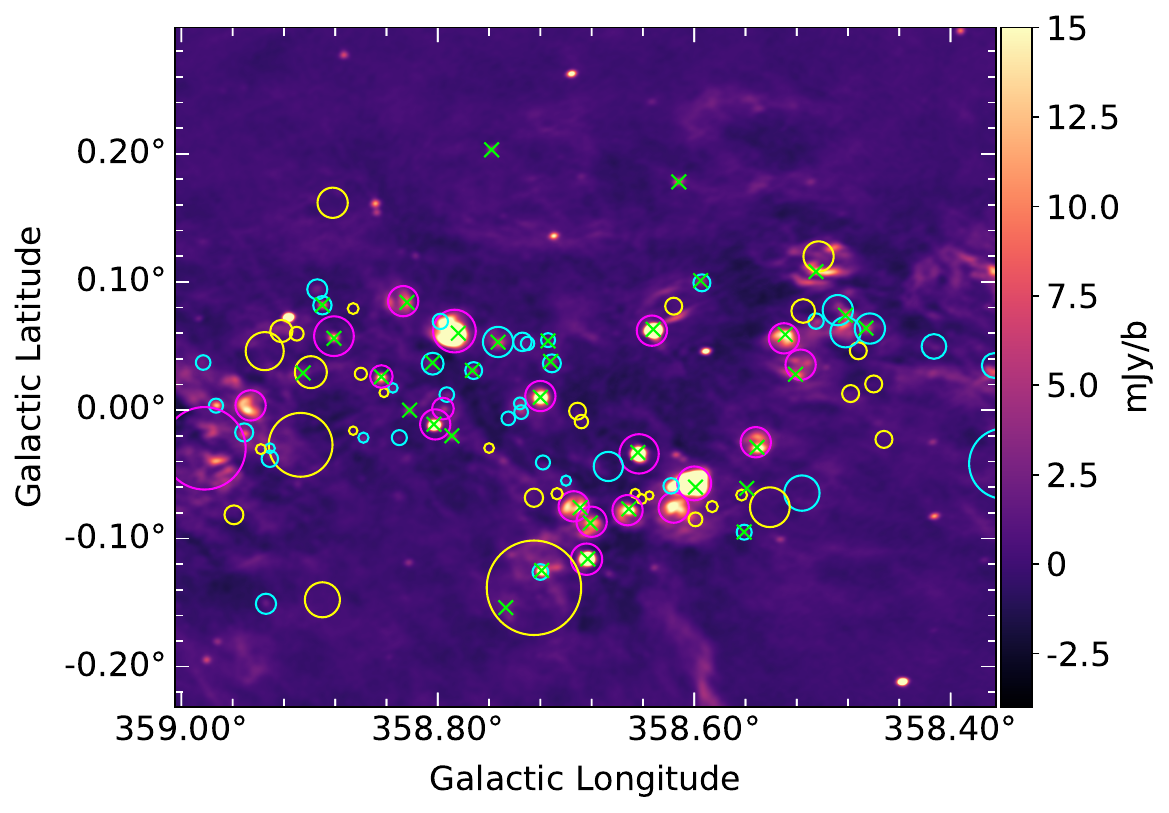}
        \caption{\ion{H}{ii} regions marked on top of the GLOSTAR VLA-D image of the target region. The magenta, cyan and gold circles represent the known, candidate and radio-quiet \ion{H}{ii} regions from the WISE \ion{H}{ii} catalog. The green crosses mark the peak flux  position of the radio sources extracted from this image that are proposed to be \ion{H}{ii} regions or candidates in this work.}
        \label{anderson}
    \end{figure}

    \begin{landscape}
    \begin{table}
        \caption{Counterparts of the extracted radio sources at different wavelengths.}
        \resizebox{\linewidth}{!}{\begin{tabular}{c c c c c c c c c c c c c c c c c c c c}
        \hline
        ID & Name& Known& Candidate & Radio-quiet& 2MASS & \multicolumn{3}{c}{GLIMPSE} & \multicolumn{2}{c}{WISE} & MIPSGAL & \multicolumn{5}{c}{Hi-GAL} & ATLASGAL & MAGPIS & MeerKAT\\
        & & \ion{H}{ii} & \ion{H}{ii} & \ion{H}{ii} & 1.25-2.16 $\mu$m & 3.6 $\mu$m & 4.5 $\mu$m & 8.0 $\mu$m & 12 $\mu$m & 22 $\mu$m & 24 $\mu$m & 70$\mu$m & 160 $\mu$m & 250 $\mu$m & 350 $\mu$m & 500 $\mu$m & 870 $\mu$m & 6 cm & 23.52 cm\\
        \hline
        1 & G358.632+0.063 & \checkmark &   &   &   &   & \checkmark & \checkmark & \checkmark & \checkmark & \checkmark & \checkmark & \checkmark & \checkmark & \checkmark & \checkmark & \checkmark & \checkmark$^c$ & \checkmark \\
        2 & G358.917+0.072 &   &   &   & \checkmark$^c$ &   &   &   &   &   &   &   &   &   &   &   &   & \checkmark$^c$ & \checkmark \\
        3 & G358.803-0.011 & \checkmark &   &   & \checkmark$^c$ & \checkmark & \checkmark & \checkmark & \checkmark & \checkmark & \checkmark$^*$ & \checkmark & \checkmark & \checkmark & \checkmark$^c$ &   &   & \checkmark$^c$ & \checkmark \\
        4 & G358.644-0.033 & \checkmark &   &   & \checkmark$^c$ & \checkmark & \checkmark & \checkmark &   & \checkmark & \checkmark$^*$ & \checkmark & \checkmark$^c$ &   &   &   &   & \checkmark$^c$ & \checkmark \\
        5 & G358.599-0.060 & \checkmark &   &   &   & \checkmark & \checkmark & \checkmark & \checkmark & \checkmark & \checkmark$^*$ & \checkmark & \checkmark$^c$ &   &   &   &   & \checkmark$^c$ & \checkmark \\
        6 & G358.683-0.116 & \checkmark &   &   & \checkmark$^c$ & \checkmark & \checkmark & \checkmark &   & \checkmark & \checkmark$^*$ & \checkmark & \checkmark & \checkmark$^c$ & \checkmark$^c$ &   &   & \checkmark$^c$ & \checkmark \\
        7 & G358.696+0.263 &   &   &   & \checkmark$^c$ &   &   &   &   &   &   &   &   &   &   &   &   & \checkmark$^c$ & \checkmark \\
        8 & G358.784+0.060 & \checkmark &   &   &   & \checkmark & \checkmark & \checkmark & \checkmark & \checkmark & \checkmark$^*$ & \checkmark &   &   &   &   &   & \checkmark$^c$ & \checkmark \\
        9 & G358.720+0.010 & \checkmark &   &   & \checkmark$^c$ & \checkmark & \checkmark & \checkmark & \checkmark & \checkmark & \checkmark$^*$ & \checkmark & \checkmark & \checkmark & \checkmark & \checkmark & \checkmark & \checkmark$^c$ & \checkmark \\
        10 & G358.719-0.125 &   & \checkmark &   &   & \checkmark$^c$ & \checkmark$^c$ &   & \checkmark & \checkmark & \checkmark$^*$ & \checkmark & \checkmark & \checkmark & \checkmark & \checkmark & \checkmark & \checkmark$^c$ & \checkmark \\
        11 & G358.592+0.046 &   &   &   &   &   &   &   &   &   &   &   &   &   &   &   &   & \checkmark$^c$ & \checkmark \\
        12 & G358.849+0.161 &   &   &   & \checkmark$^c$ & \checkmark & \checkmark & \checkmark$^c$ &   &   &   &   &   &   &   &   &   & \checkmark$^c$ & \checkmark \\
        13 & G358.710+0.136 &   &   &   & \checkmark$^c$ & \checkmark & \checkmark & \checkmark$^c$ &   &   &   &   &   &   &   &   &   & \checkmark$^c$ & \checkmark \\
        14a & G358.529+0.059 &\checkmark & & &\checkmark$^c$ & \checkmark& \checkmark& \checkmark&\checkmark & \checkmark& \checkmark&\checkmark & \checkmark$^c$& & & & & \checkmark$^c$ &\checkmark\\
        14b & G358.521+0.028 & \checkmark & & & && & & \checkmark & \checkmark & \checkmark &\checkmark$^c$ & & & & & & &\checkmark\\
        14c & G358.482+0.074 & & \checkmark & & \checkmark$^c$ & & & \checkmark& \checkmark& \checkmark& \checkmark& \checkmark& & & & & & &\checkmark\\
        15 & G358.551-0.029 & \checkmark &   &   &   &   &   & \checkmark & \checkmark & \checkmark & \checkmark & \checkmark$^c$ &   &   &   &   &   &   & \checkmark \\
        16 & G358.651-0.077 & \checkmark &   &   & \checkmark$^c$ & \checkmark & \checkmark & \checkmark &   & \checkmark & \checkmark$^*$ & \checkmark &   &   &   &   &   &   & \checkmark \\
        17 & G358.844+0.026 & \checkmark &   &   & \checkmark$^c$ & \checkmark & \checkmark & \checkmark & \checkmark & \checkmark & \checkmark$^*$ & \checkmark & \checkmark$^c$ &   &   &   &   & \checkmark$^c$ & \checkmark \\
        18 & G358.881+0.056 & \checkmark &   &   & \checkmark$^c$ & \checkmark & \checkmark & \checkmark & \checkmark & \checkmark & \checkmark$^*$ & \checkmark & \checkmark & \checkmark & \checkmark$^c$ &   & \checkmark & \checkmark$^c$ & \checkmark \\
        19 & G358.505+0.108 &   &   & \checkmark  &   &   &   &   & \checkmark & \checkmark &   &   &   &   &   &   &   &   & \checkmark \\
        20 & G358.890+0.082 &   & \checkmark &   & \checkmark$^c$ & \checkmark & \checkmark & \checkmark & \checkmark & \checkmark & \checkmark & \checkmark & \checkmark & \checkmark & \checkmark$^c$ &   & \checkmark &   & \checkmark \\
        21a & G358.689–0.076 & \checkmark & & & \checkmark$^c$ &\checkmark&\checkmark & \checkmark& \checkmark& \checkmark& \checkmark$^*$& \checkmark& & & & & & &\checkmark\\
        21b & G358.681–0.088 & \checkmark & & & &\checkmark&\checkmark &\checkmark & & & \checkmark$^*$&\checkmark & & & & & & &\checkmark\\
        22 & G358.612+0.178 &   &   &   &   &   &   &   & \checkmark & \checkmark & \checkmark & \checkmark$^c$ &   &   & \checkmark$^c$ &   &   & \checkmark$^c$ & \checkmark \\
        23 & G358.758+0.203 &   &   &   & \checkmark$^c$ & \checkmark & \checkmark & \checkmark & \checkmark & \checkmark & \checkmark & \checkmark$^c$ &   &   &   &   &   & \checkmark$^c$ & \checkmark \\
        24 & G358.824+0.084 & \checkmark &   &   & \checkmark$^c$ & \checkmark & \checkmark & \checkmark & \checkmark & \checkmark & \checkmark & \checkmark & \checkmark$^c$ &   &   &   &   &   & \checkmark \\
        25 & G358.753+0.053 &   & \checkmark &   &   &   & \checkmark & \checkmark & \checkmark & \checkmark & \checkmark & \checkmark &   &   &   &   &   &   & \checkmark \\
        26 & G358.789-0.020 &   &   &   & \checkmark$^c$ & \checkmark$^c$ & \checkmark$^c$ & \checkmark$^c$ & \checkmark & \checkmark & \checkmark & \checkmark$^c$ &   &   &   &   &   &   & \checkmark \\
        27 & G358.614-0.034 &   &   &   & \checkmark$^c$ &   &   &   & \checkmark & \checkmark &   &   &   &   &   &   &   &   & \checkmark \\
        28 & G358.823-0.118 &   &   &   &   & \checkmark$^c$ & \checkmark$^c$ &   & \checkmark & \checkmark &   & \checkmark$^c$ &   &   &   &   &   &   & \checkmark \\
        29 & G358.822+0.000 &   &   &   & \checkmark$^c$ &   &   &   & \checkmark & \checkmark & \checkmark & \checkmark$^c$ &   &   &   &   &   & \checkmark$^c$ & \checkmark \\
        30 & G358.773+0.031 &   & \checkmark &   & \checkmark$^c$ &   &   & \checkmark & \checkmark & \checkmark & \checkmark & \checkmark & \checkmark & \checkmark &   &   &   &   & \checkmark \\
        31 & G358.595+0.101 &   & \checkmark &   &   &   &   & \checkmark & \checkmark & \checkmark & \checkmark & \checkmark$^c$ &   &   &   &   &   &   & \checkmark \\
        32 & G358.533-0.100 &   &   &   & \checkmark$^c$ & \checkmark & \checkmark &   & \checkmark & \checkmark & \checkmark &   &   &   &   &   &   &   & \checkmark \\
        33 & G358.466+0.064 &   & \checkmark &   & \checkmark$^c$ & \checkmark & \checkmark & \checkmark & \checkmark & \checkmark & \checkmark & \checkmark & \checkmark$^c$ &   &   &   &   &   & \checkmark \\
        34 & G358.712+0.038 &   & \checkmark &   & \checkmark$^c$ & \checkmark & \checkmark & \checkmark & \checkmark & \checkmark & \checkmark & \checkmark &   &   &   &   &   &   & \checkmark \\
        35 & G358.747-0.154 &   &   &  \checkmark & \checkmark$^c$ & \checkmark & \checkmark & \checkmark & \checkmark & \checkmark & \checkmark$^c$ &   &   &   &   &   &   &   &   \\
        36 & G358.622+0.093 &   &   &   &   & \checkmark$^c$ & \checkmark$^c$ &   &   &   & \checkmark &   &   &   &   &   &   &   &   \\
        37 & G358.905+0.029 &   &   &  \checkmark &   & \checkmark$^c$ & \checkmark$^c$ &   &   &   &   &   &   &   &   &   &   &   & \checkmark \\
        38 & G358.557+0.243 &   &   &   &   & \checkmark$^c$ & \checkmark$^c$ &   & \checkmark & \checkmark &   &   &   &   &   &   &   &   &   \\
        39 & G358.510+0.178 &   &   &   & \checkmark$^c$ & \checkmark & \checkmark & \checkmark$^c$ & \checkmark & \checkmark &  &   &   &   &   &   &   &   & \checkmark \\
        40 & G358.561-0.095 &   & \checkmark &   & \checkmark$^c$ & \checkmark & \checkmark & \checkmark & \checkmark & \checkmark & \checkmark & \checkmark &   &   &   &   &   &   & \checkmark \\
        41 & G358.714+0.054 &   & \checkmark &   &   & \checkmark & \checkmark & \checkmark & \checkmark & \checkmark & \checkmark & \checkmark$^c$ &   &   &   &   &   &   & \checkmark \\
        42 & G358.631-0.122 &   &   &   & \checkmark$^c$ & \checkmark$^c$ & \checkmark$^c$ &   &   &   &   &   &   &   &   &   &   &   & \checkmark \\
        43 & G358.804+0.037 &   & \checkmark &   &   &   & \checkmark & \checkmark & \checkmark & \checkmark & \checkmark & \checkmark &   &   &   &   &   &   & \checkmark \\
        44 & G358.633+0.240 &   &   &   & \checkmark$^c$ &   &   &   & \checkmark & \checkmark &   & \checkmark$^c$ &   &   &   &   &   &   &   \\
        45 & G358.559-0.061 &   &   &   & \checkmark$^c$ & \checkmark$^c$ & \checkmark$^c$ &   & \checkmark & \checkmark &   &   &   &   &   & \checkmark$^c$ &   &   & \checkmark \\
        46 & G358.688+0.182 &   &   &   & \checkmark$^c$ & \checkmark$^c$ & \checkmark$^c$ & \checkmark$^c$ & \checkmark & \checkmark &   &   &   &   &   &   &   &   &   \\
        \hline
        \end{tabular}}
        \label{count}
        \tablebib{The known, candidate and radio-quiet \ion{H}{ii} region columns list the cross-matches from the WISE \ion{H}{ii} catalog \citep[V2.3;][]{2014Anderson} with a tickmark (\checkmark). The tickmarks in the remaining columns list the source with observed signature in the respective images. The superscript (*) represents the saturation in the MIPSGAL 24 $\mu$m image. The sources having counterparts in the respective catalogs for GLIMPSE 3.6, 4.5 and 8 $\mu$m, MIPSGAL 24 $\mu$m, and Hi-GAL 70-500 $\mu$m emission, given in Tab.~\ref{tab:numberofcounterparts}, but not displaying any clear visual signature in the images are represented with the superscript (\textsuperscript{c}). The counterparts for 2MASS 1.25-2.16 $\mu$m and MAGPIS 6 cm are determined only from their respective catalogs provided in Tab.~\ref{tab:numberofcounterparts}, and are also marked with the superscript (\textsuperscript{c}). The MeerKAT counterparts were visually identified from the `1.28 GHz MeerKAT GC mosaic' \citep{2022meerkat}.}
    \end{table}
    \end{landscape}

    \begin{table*}[h!]
        \centering
        \caption{Catalogs used to identify counterparts from the NIR to the radio.}
        \resizebox{\linewidth}{!}{\begin{tabular}{c c c c c }
        \hline
        Survey & Bands & Catalog & Angular  & Search  \\   
        & ($\lambda$) &  & Resolution & Radius  \\
        \hline
        \multirow{3}{*}{2MASS} & 1.25 $\mu$m & 2-MASS All-Sky Point Source Catalog & \multirow{3}{*}{$\thicksim$4$''$} & \multirow{3}{*}{4$''$} \\
        & 1.65 $\mu$m & \citep[PSC;][]{2003twomass}& &  \\
        & 2.16 $\mu$m & & &   \\
        \hline
        \multirow{4}{*}{GLIMPSE} & 3.6 $\mu$m &  & \multirow{4}{*}{$\thicksim$2$''$} & \multirow{4}{*}{2$''$}  \\
        & 4.5 $\mu$m & GLIMPSE II Epoch 1 December '08 Catalog & &  \\
        & 5.8 $\mu$m &\citep{2020ipac.data.I227G} & &  \\
        & 8.0 $\mu$m & & &  \\
        \hline
        \multirow{4}{*}{WISE } & 3.4 $\mu$m  &  &6.1$''$& \multirow{4}{*}{10$''$} \\
        & 4.6 $\mu$m & WISE All-Sky Source Catalog &6.4$''$ & \\
        & 12 $\mu$m &  \citep{2020Wise}& 6.5$''$&  \\
        & 22 $\mu$m & & 12$''$&   \\
        \hline
        \multirow{2}{*}{MIPSGAL} & \multirow{2}{*}{24$\mu$m}  & MIPSGAL Catalog & \multirow{2}{*}{6$''$}& \multirow{2}{*}{6$''$} \\
         & & \citep{2020mipsgal} && \\
        \hline
        \multirow{5}{*}{Hi--GAL}& 70$\mu$m & Hi-GAL 70 micron Photometric Catalog \citep{2020ipac.data..I27H} & 6$''$ & \multirow{3}{*}{10$''$}  \\
        & 160 $\mu$m &Hi-GAL 160 micron Photometric Catalog \citep{2020ipac.data..I25H}& 12$''$ & \\
        & 250 $\mu$m & Hi-GAL 250 micron Photometric Catalog \citep{2020ipac.data..I24H}& 18$''$ & \\
        & 350 $\mu$m & Hi-GAL 350 micron Photometric Catalog \citep{2020ipac.data..I28H} & 24$''$ & \\
        & 500 $\mu$m & Hi-GAL 500 micron Photometric Catalog \citep{2020ipac.data..I26H} & 35$''$ & \\
        \hline
        \multirow{2}{*}{ATLASGAL}  & \multirow{2}{*}{870 $\mu$m} & ATLASGAL - Complete Compact Source & \multirow{2}{*}{19.2$''$} & \multirow{2}{*}{18$''$}   \\
        && Catalogue \citep[CSC;][]{2014urquhart} &&\\
        \hline
        \multirow{2}{*}{MAGPIS}  & \multirow{2}{*}{6 cm} & New catalogs of compact radio sources & \multirow{2}{*}{6$''$} & \multirow{2}{*}{6$''$}   \\
         &&in the Galactic plane \citep{2005magpis} & &\\
        \hline
        
        \end{tabular}}
        \label{tab:numberofcounterparts}
        \tablebib{The infrared counterparts were searched using NASA/IPAC Infrared Science Archive (\url{https://irsa.ipac.caltech.edu/frontpage/}), and the submillimeter (ATLASGAL) and radio (MAGPIS) counterparts were searched using catalogs mentioned in the table. The respective angular resolution and the search radius used to identify counterparts are listed in the last two columns.}
    \end{table*}
    \vspace{5em}

    \begin{table*}[h]
        \centering
        \caption{Classification scheme of all the radio sources toward G358.69+0.03.}
        \resizebox{\linewidth}{!}{\begin{tabular}{c c c c c c c c c c c c c }
        \hline
        ID & Name &  Glostar & \multicolumn{2}{c}{Spitzer} & Hi-GAL & ATLASGAL& Cold dust &WISE & Known  & Candidate & Radio-quiet & Final\\
         & & in-band $\alpha$& 8 $\mu$m & 24$\mu$m & 70$\mu$m & 870 $\mu$m & Clumps & Colors & \ion{H}{ii} & \ion{H}{ii}& \ion{H}{ii} & Classification \\
        \hline
        1 & G358.632+0.063 & -0.25$\pm$0.01 & \checkmark & \checkmark & \checkmark & \checkmark & \checkmark & \checkmark & \checkmark &   &  & \ion{H}{ii} \\
        2 & G358.917+0.072 & -1.59$\pm$0.02 &   &   &   &   &   &   &   &   &  & cEG \\
        3 & G358.803-0.011 & -0.26$\pm$0.03 & \checkmark & \checkmark$^*$ & \checkmark &   & \checkmark & \checkmark & \checkmark &   &  & \ion{H}{ii} \\
        4 & G358.644-0.033 & -0.28$\pm$0.02 & \checkmark & \checkmark$^*$ & \checkmark &   &   &   & \checkmark &   &  & \ion{H}{ii} \\
        5 & G358.599-0.060  & - & \checkmark & \checkmark$^*$ & \checkmark &   &   & \checkmark & \checkmark &   &  &  \ion{H}{ii}\\
        6 & G358.683-0.116 & -0.32$\pm$0.03 & \checkmark & \checkmark$^*$ & \checkmark &   & \checkmark &   & \checkmark &   &  & \ion{H}{ii} \\
        7 & G358.696+0.263 & - &   &   &   &   &   &   &   &   &  &  cEG\\
        8 & G358.784+0.060 & - & \checkmark & \checkmark$^*$ & \checkmark &   &   & \checkmark & \checkmark &   &  & \ion{H}{ii} \\
        9 & G358.720+0.010 & - & \checkmark & \checkmark$^*$ & \checkmark & \checkmark & \checkmark & \checkmark & \checkmark &   &  & \ion{H}{ii} \\
        10 & G358.719-0.125  & - &   & \checkmark$^*$ & \checkmark & \checkmark & \checkmark & \checkmark &   & \checkmark &  &  \ion{H}{ii}\\
        11 & G358.592+0.046  & -1.33$\pm$0.06 &   &   &   &   &   &  &   &   &  &  cEG\\
        12 & G358.849+0.161 & - & \checkmark$^c$ &   &   &   &   &   &   &   &  &  cEG\\
        13 & G358.710+0.136 & - & \checkmark$^c$ &   &   &   &   &   &   &   &  & cEG \\
        14a & G358.529+0.059  & -& \checkmark &\checkmark &\checkmark & & & \checkmark&\checkmark &&&\ion{H}{ii}\\
        14b & G358.521+0.028 & -& &\checkmark &\checkmark$^c$ & & & \checkmark &\checkmark&&&\ion{H}{ii}\\
        14c & G358.482+0.074 & -& \checkmark &\checkmark &\checkmark & & &\checkmark & &\checkmark&&\ion{H}{ii}\\
        15 & G358.551-0.029 & - & \checkmark & \checkmark & \checkmark$^c$ &   &   & \checkmark & \checkmark &   &  & \ion{H}{ii} \\
        16 & G358.651-0.077  & - & \checkmark & \checkmark$^*$ & \checkmark &   &   &   & \checkmark &   &  & \ion{H}{ii} \\
        17 & G358.844+0.026  & -0.34$\pm$0.07 & \checkmark & \checkmark$^*$ & \checkmark &   &   & \checkmark & \checkmark &   &  & \ion{H}{ii} \\
        18 & G358.881+0.056  & -0.13$\pm$0.08 & \checkmark & \checkmark$^*$ & \checkmark & \checkmark & \checkmark & \checkmark & \checkmark &   &  &  \ion{H}{ii}\\
        19 & G358.505+0.108 & - &   &   &   &   &   & \checkmark &   &   & \checkmark &  c\ion{H}{ii}\\
        20 & G358.890+0.082 & -0.33$\pm$0.1 & \checkmark & \checkmark & \checkmark & \checkmark & \checkmark & \checkmark &   & \checkmark &  &  \ion{H}{ii}\\
        21a & G358.689–0.076 & - &\checkmark &\checkmark$^*$ &\checkmark & & & \checkmark &\checkmark&&&\ion{H}{ii}\\
        21b & G358.681-0.088  & - &\checkmark &\checkmark$^*$ &\checkmark & & & &\checkmark&&&\ion{H}{ii}\\
        22 & G358.612+0.178  & 0.02$\pm$0.13 &   & \checkmark & \checkmark$^c$ &   & \checkmark & \checkmark &   &   &  & c\ion{H}{ii} \\
        23 & G358.758+0.203  & -0.34$\pm$0.15 & \checkmark$^c$ & \checkmark & \checkmark$^c$ &   &   & \checkmark &   &   &  & c\ion{H}{ii} \\
        24 & G358.824+0.084  & - & \checkmark & \checkmark & \checkmark &   &   & \checkmark & \checkmark &   &  &  \ion{H}{ii}\\
        25 & G358.753+0.053  & - & \checkmark & \checkmark & \checkmark &   &   & \checkmark &   & \checkmark &  & \ion{H}{ii} \\
        26 & G358.789-0.020  & 0.28$\pm$0.28 & \checkmark$^c$ & \checkmark & \checkmark$^c$ &   &   & \checkmark &   &   &  & c\ion{H}{ii} \\
        27 & G358.614-0.034  & -0.89$\pm$0.41 &   &   &   &   &   & \checkmark &   &   &  & Rad \\
        28 & G358.823-0.118  & -0.1$\pm$0.17 &   &   & \checkmark$^c$ &   &   & $\times$ &   &   &  & cPNe \\
        29 & G358.822+0.000  & -0.09$\pm$0.23 &   & \checkmark & \checkmark$^c$ &   &   & \checkmark &   &   &  & c\ion{H}{ii} \\
        30 & G358.773+0.031  & -0.17$\pm$0.25 & \checkmark & \checkmark & \checkmark &   & \checkmark & \checkmark &   & \checkmark &  & \ion{H}{ii} \\
        31 & G358.595+0.101  & 0.13$\pm$0.38 & \checkmark & \checkmark & \checkmark$^c$ &   &   & \checkmark &   & \checkmark &  &  \ion{H}{ii}\\
        32 & G358.533-0.100  & -0.99$\pm$0.32 &   & \checkmark &   &   &   & \checkmark &   &   &  & cPNe \\
        33 & G358.466+0.064  & - & \checkmark & \checkmark & \checkmark &   &   & \checkmark &   & \checkmark &  & \ion{H}{ii} \\
        34 & G358.712+0.038  & - & \checkmark & \checkmark & \checkmark &   &   & \checkmark &   & \checkmark &  & \ion{H}{ii} \\
        35 & G358.747-0.154  & - & \checkmark & \checkmark$^c$ &   &   &   & \checkmark &   &   & \checkmark &  c\ion{H}{ii}\\
        36 & G358.622+0.093  & - &   & \checkmark &   &   &   &   &   &   &  & Rad \\
        37 & G358.905+0.029   & - &   &   &   &   &   &   &   &   & \checkmark & c\ion{H}{ii} \\
        38 & G358.557+0.243   & - &   &   &   &   &   & \checkmark &   &   &  & Rad \\
        39 & G358.510+0.178  & - & \checkmark$^c$ &  &   &   &   & $\times$ &   &   &  & Rad \\
        40 & G358.561-0.095  & - & \checkmark & \checkmark & \checkmark &   &   & \checkmark &   & \checkmark &  &  \ion{H}{ii}\\
        41 & G358.714+0.054  & - & \checkmark & \checkmark & \checkmark$^c$ &   &   & \checkmark &   & \checkmark &  &  \ion{H}{ii}\\
        42 & G358.631-0.122  & - &   &   &   &   &   &   &   &   &  &  cEG\\
        43 & G358.804+0.037  & - & \checkmark & \checkmark & \checkmark &   &   & \checkmark &   & \checkmark &  &  \ion{H}{ii}\\
        44 & G358.633+0.240  & - &   &   & \checkmark$^c$ &   &   & $\times$ &   &   &  &  cPNe\\
        45 & G358.559-0.061  & - &   &   &   &   & \checkmark & \checkmark &   &   &  &  c\ion{H}{ii}\\
        46 & G358.688+0.182  & - & \checkmark$^c$ &   &   &   &   & $\times$ &   &   &  & cPNe \\
        \hline
        \end{tabular}}
        \label{hiianalysis}
        \tablebib{The table provides the source ID, name, and the GLOSTAR in-band spectral index. The observed counterparts in the GLIMPSE 8 $\mu$m, MIPSGAL 24 $\mu$m, Hi-GAL 70 $\mu$m, and the ATLASGAL 870 $\mu$m images are marked with a tick (\checkmark). The superscript (*) represents the saturation in the 24 $\mu$m MIPSGAL images. The sources having counterparts in the respective catalogs for 8, 24 and 70 $\mu$m emission, given in Tab.~\ref{tab:numberofcounterparts}, but not displaying any clear visual signature in the images are represented with the superscript (\textsuperscript{c}). The `Cold dust clumps' column displays the corresponding cross-matches in either the Hi-GAL 250, 350, and 500 $\mu$m bands or the ATLASGAL 870 $\mu$m band. The sources occupying the same parameter space in the MIR color-magnitude plot where all of \ion{H}{ii} regions and a vast majority of the PNe are located, as proposed by \cite{2019Sac}, are marked with a tick under the column `WISE colors', and the ones not present in the same parameter space are marked with a cross ($\times$). The next three columns list the sources with cross-matches with the known, candidate and radio-quiet \ion{H}{ii} regions from the WISE \ion{H}{ii} catalog \citep[V2.3;][]{2014Anderson}. The final column represents the classification of these sources as \ion{H}{ii} regions, and the candidates of \ion{H}{ii} regions (c\ion{H}{ii}), Planetary nebulae (cPNe) and Extragalactic sources (cEG). `Rad' represents the unidentified radio sources.}
    \end{table*}

    \begin{table*}
        \caption{Physical properties of the radio sources not classified as \ion{H}{ii} regions.} 
        \label{maintable}
        \resizebox{\linewidth}{!}{\begin{tabular}{c c c c c c c c c c}
        \hline
        ID & $n_{\mathrm{pix}}$ & \textit{l} & \textit{b} & Name &$\sigma_{\mathrm{noise}}^p$ & SNR & S$^{\mathrm{p}}_{\mathrm{OBS}}$ & $F_{\mathrm{5.8 GHz}}$  &Y-factor \\
        &  & [\textdegree] & [\textdegree] & & [mJy/b] & & [mJy/b] & [mJy] &  \\
        \hline
        2 & 746 & 358.917 & +0.072 & G358.917+0.072 & 0.467 & 149.0 & 69.10 & 87.10 & 1.26  \\
        7 & 215 & 358.696 & +0.263 & G358.696+0.263 & 0.194 & 109.0 & 21.10 & 19.60 & 0.93 \\
        11 & 126 & 358.592 & +0.046 & G358.592+0.046 & 0.448 & 53.9 & 23.90 & 18.10 & 0.76\\
        12 & 272 & 358.849 & +0.161 & G358.849+0.161 & 0.293 & 39.8 & 11.70 & 17.10 & 1.46\\
        13 & 281 & 358.710 & +0.136 & G358.710+0.136 & 0.428 & 36.4 & 15.50 & 19.20 & 1.24\\
        27 & 165 & 358.614 & --0.034 & G358.614--0.034 & 0.553 & 12.2 & 6.69 & 8.97 & 1.34\\
        28 &  75 & 358.823 & --0.118 & G358.823--0.118 & 0.290 & 11.8 & 3.36 & 2.42 &  0.72\\
        32 & 96 & 358.533 & --0.100 & G358.553--0.100 & 0.416 & 10.3 & 4.27 & 3.89 & 0.91\\
        36 & 403 & 358.622 & +0.093 & G358.622+0.093 & 0.433 & 7.6 & 3.56 & 13.70 & 3.85\\
        38 & 1382 & 358.557 & +0.243 & G358.557+0.243 & 0.345 & 5.1 & 2.37 & 35.20 & 14.85\\
        39 & 89   & 358.510 & +0.178 & G358.510+0.178 & 0.450 & 6.1 & 2.85 & 2.97 & 1.04\\
        42 & 43 & 358.631 & --0.122 & G358.631--0.122 & 0.613 & 5.5 & 3.36 & 1.82 & 0.54\\
        44 & 44 & 358.633 & +0.240 & G358.633+0.240 & 0.342 & 5.4 & 1.82 & 1.05 & 0.58\\
        46 & 54 & 358.688 & +0.182 & G358.688+0.182 & 0.360 & 5.1 & 1.81 & 1.27 & 0.70\\
        \hline
    \end{tabular}}
    \label{propertiesnonhii}
    \tablebib{The columns represent the source ID, number of pixels ($n_{\mathrm{pix}}$), Galactic coordinates (\textit{l}, \textit{b}), the source name, rms noise at position of peak pixel ($\sigma^p_{\mathrm{noise}}$), observed SNR, peak surface brightness (S$^{\mathrm{p}}_{\mathrm{OBS}}$), flux density derived using the VLA-D 5.8 GHz radio continuum data and the Y-factor.}
    \end{table*}

    \begin{landscape}
    \begin{table}
        \caption{Physical properties of the \ion{H}{ii} regions.} 
        \label{maintable}
        \resizebox{\linewidth}{!}{\begin{tabular}{c c c c c c c c c c c c c c c c c}
        \hline
        ID & $n_{\mathrm{pix}}$ & \textit{l} & \textit{b} & Name &$\sigma_{\mathrm{noise}}^p$ & SNR & S$^{\mathrm{p}}_{\mathrm{OBS}}$ & $F_{\mathrm{5.8 GHz}}$  &Y-factor &  log$_{\mathrm{10}}$($N_{\mathrm{ly}}$) & Sp. & $R_{\mathrm{\ion{H}{ii}}}$ & $t_{\mathrm{dyn}}$ & $n_e$ & EM & $\alpha$  \\ 
        &  & [\textdegree] & [\textdegree] & & [mJy/b] & & [mJy/b] & [mJy] &  & log$_{\mathrm{10}}$[s$^{\mathrm{-1}}$] & Type  & [pc] & [Myr]  & [cm$^{-3}$] & [pc cm$^{-6}$] &  \\
        \hline
        1 & 1137 & 358.632 & +0.063 & G358.632+0.063 & 0.429 & 552.0 & 236.00 & 401.00 & 1.70 & 48.49 & O7.5  & 1.89 & 0.31 & 97.17 & 35705.9 &--0.25 $\pm$ 0.01\\
        3 & 591 & 358.803 & --0.011 & G358.803--0.011 & 0.528 & 144.0 & 75.40 & 107.00 & 1.42 & 47.91 & O9  & 1.36 & 0.25 & 82.00 & 18329.6 &--0.26 $\pm$ 0.03\\
        4 & 433 & 358.644 & --0.033 & G358.644--0.033 & 0.631 & 124.0 & 77.70 & 145.00 & 1.87 & 48.04 & O9  & 1.17 & 0.17 & 120.53 & 33902.9 &--0.28 $\pm$ 0.02 \\
        5 & 4757 & 358.599 & --0.060 & G358.599--0.060 & 0.592 & 122.0 & 71.80 & 872.00 & 12.14 & 48.82 & O6.5  & 3.87 & 0.94 & 48.98 & 18558.4 & -- \\
        6 & 459 & 358.683 & --0.116 & G358.683--0.116 & 0.717 & 116.0 & 82.70 & 160.00 & 1.93 & 48.09 & O9  & 1.20 & 0.17 & 121.20 & 35291.0 &--0.32 $\pm$ 0.03\\
        8 & 1875 & 358.784 & +0.060 & G358.784+0.060 & 0.568 & 98.7 & 55.90 & 434.00 & 7.76 & 48.52 & O7.5  & 2.43 & 0.49 & 69.47 & 23433.9 &-- \\
        9 & 1086 & 358.720 & +0.010 & G358.720+0.010 & 0.554 & 69.4 & 38.20 & 128.00 & 3.35 & 47.99 & O9  & 1.85 & 0.41 & 56.82 & 11932.6 &-- \\
        10 & 2869 & 358.719 & --0.125 & G358.719--0.125 & 0.477 & 58.9 & 28.10 & 154.00 & 5.48 & 48.07 & O9  & 3.00 & 0.94 & 30.08 & 5434.3 &-- \\
        14a & 1239 & 358.529 & +0.059 & G358.529+0.059 & 0.559 & 35.1 & 19.58 & 133.29 & 6.81 & 48.01& O9 & 1.97&0.46 &52.47 &10867.7 & --\\
        14b & 794 & 358.521 & +0.028 & G358.521+0.028 & 0.616 & 17.5 & 10.76 & 66.88 & 6.22 & 47.71& O9.5&1.58 &0.37 &51.96 & 8530.3 & --\\
        14c & 602 & 358.482 & +0.074 & G358.482+0.074 & 0.784 & 12.9 & 10.46 & 58.58 & 5.60 & 47.65& O9.5& 1.38 & 0.30 &59.85 &9855.0 & --\\
        15 & 1081 & 358.551 & --0.029 & G358.551--0.029 & 0.559 & 34.6 & 19.30 & 108.00 & 5.60 & 47.92 & O9  & 1.84 & 0.43 & 52.38 & 10114.7 & -- \\
        16 & 796 & 358.651 & --0.077 & G358.651--0.077 & 0.705 & 34.7 & 24.20 & 90.20 & 3.73 & 47.84 & O9.5  & 1.58 & 0.34 & 60.22 & 11472.3 & -- \\
        17 & 260 & 358.844 & +0.026 & G358.844+0.026 & 0.564 & 34.8 & 19.30 & 29.70 & 1.54 & 47.36 & B0.5  & 0.90 & 0.17 & 79.97 & 11564.8 &--0.34 $\pm$ 0.07\\
        18 & 195 & 358.881 & +0.056 & G358.881+0.056 & 0.503 & 31.6 & 15.70 & 19.90 & 1.27 & 47.18 & B0.5  & 0.78 & 0.14 & 81.23 & 10331.8 &--0.13 $\pm$ 0.08\\
        19 & 2491 & 358.505 & +0.108 & G358.505+0.108 & 0.555 & 29.6 & 16.40 & 180.00 & 10.98 & 48.14 & O8.5  & 2.80 & 0.80 & 36.15 & 7315.7 &-- \\
        20 & 139 & 358.890 & +0.082 & G358.890+0.082 & 0.463 & 26.3 & 12.20 & 12.00 & 0.98 & 46.96 & B1  & 0.66 & 0.12 & 81.31 & 8740.2 &--0.33 $\pm$ 0.10\\
        21a & 557 & 358.689 & --0.076 & G358.689--0.076 & 0.734 & 23.3 & 17.13 & 93.24 & 5.44 &47.85 &O9.5 & 1.32& 0.24& 80.00& 16940.2 & --\\
        21b & 389 & 358.681 & --0.088 & G358.681--0.088 & 0.733 & 18.0 & 13.17 & 55.43 & 4.21 & 47.63 &O9.5 & 1.11&0.20 &80.74 & 14418.4& --\\
        22 & 160 & 358.612 & +0.178 & G358.612+0.178 & 0.442 & 22.1 & 9.65 & 10.90 & 1.13 & 46.92 & B1  & 0.71 & 0.14 & 69.73 & 6897.0 &+0.02 $\pm$ 0.13\\
        23 & 154 & 358.758 & +0.203 & G358.758+0.203 & 0.253 & 21.0 & 5.30 & 5.55 & 1.05 & 46.63 & B1  & 0.70 & 0.16 & 51.20 & 3648.6 &--0.34 $\pm$ 0.15\\
        24 & 1150 & 358.824 & +0.084 & G358.824+0.084 & 0.529 & 19.6 & 10.30 & 77.50 & 7.52 & 47.77 & O9.5  & 1.90 & 0.50 & 42.36 & 6822.8 & -- \\
        25 & 451 & 358.753 & +0.053 & G358.753+0.053 & 0.539 & 15.2 & 8.17 & 24.90 & 3.05 & 47.28 & B0.5  & 1.19 & 0.29 & 48.45 & 5589.6 & -- \\
        26 & 111 & 358.789 & --0.020 & G358.789--0.020 & 0.535 & 13.0 & 6.91 & 6.56 & 0.95 & 46.70 & B1  & 0.59 & 0.12 & 71.16 & 5983.3 & +0.28 $\pm$ 0.28\\
        29 & 86 & 358.822 & +0.000 & G358.822+0.000 & 0.539 & 11.6 & 6.23 & 5.02 & 0.81 & 46.58 & B1  & 0.52 & 0.10 & 75.38 & 5909.7 & --0.09 $\pm$ 0.23\\
        30 & 98 & 358.773 & +0.031 & G358.773+0.031 & 0.554 & 10.7 & 5.87 & 5.68 & 0.97 & 46.64 & B1  & 0.55 & 0.11 & 72.70 & 5867.8 &--0.17 $\pm$ 0.25 \\
        31 & 184 & 358.595 & +0.101 & G358.595+0.101 & 0.431 & 10.6 & 4.52 & 7.81 & 1.73 & 46.78 & B1  & 0.76 & 0.18 & 53.15 & 4297.2 & +0.13 $\pm$ 0.38\\
        33 & 168 & 358.466 & +0.064 & G358.466+0.064 & 0.848 & 9.2 & 7.78 & 12.60 & 1.62 & 46.98 & B1  & 0.73 & 0.14 & 72.28 & 7593.1 & -- \\
        34 & 135 & 358.712 & +0.038 & G358.712+0.038 & 0.506 & 8.6 & 4.37 & 6.10 & 1.40 & 46.67 & B1  & 0.65 & 0.14 & 59.25 & 4574.6 & -- \\
        35 & 733 & 358.747 & --0.154 & G358.747--0.154 & 0.301 & 7.2 & 2.52 & 19.10 & 7.58 & 47.16 & B0.5  & 1.52 & 0.48 & 29.48 & 2638.1 &-- \\
        37 & 236 & 358.905 & +0.029 & G358.905+0.029 & 0.545 & 6.2 & 3.79 & 9.63 & 2.54 & 46.87 & B1  & 0.86 & 0.21 & 48.97 & 4131.1 & -- \\
        40 & 38 & 358.561 & --0.095 & G358.561--0.095 & 0.556 & 5.8 & 3.20 & 1.54 & 0.48 & 46.07 & >B1.5  & 0.35 & 0.07 & 77.04 & 4102.9 & -- \\
        41 & 54 & 358.714 & +0.054 & G358.714+0.054 & 0.499 & 5.6 & 2.77 & 1.87 & 0.68 & 46.15 & B1.5  & 0.41 & 0.09 & 65.23 & 3505.9 & -- \\
        43 & 51 & 358.804 & +0.037 & G358.804+0.037 & 0.574 & 5.4 & 3.09 & 2.01 & 0.65 & 46.19 & B1.5  & 0.40 & 0.08 & 70.59 & 3990.1 & -- \\
        45 & 70 & 358.559 & --0.061 & G358.559--0.061 & 0.572 & 5.2 & 2.97 & 2.71 & 0.91 & 46.32 & B1.5  & 0.47 & 0.10 & 64.63 & 3919.5 & -- \\
        \hline
    \end{tabular}}
    \label{propertieshii}
    \tablebib{The columns represent the source ID, number of pixels ($n_{\mathrm{pix}}$), Galactic coordinates (\textit{l}, \textit{b}), the source name, rms noise at position of peak pixel ($\sigma^p_{\mathrm{noise}}$), observed SNR, peak surface brightness (S$^{\mathrm{p}}_{\mathrm{OBS}}$), flux density derived using the VLA-D 5.8 GHz radio continuum data, Y-factor, Lyman continuum photon rate ($N_{\mathrm{ly}}$), the spectral type, effective radii of the \ion{H}{ii} regions ($R_{\mathrm{\ion{H}{ii}}}$), the dynamical age ($t_{\mathrm{dyn}}$), electron number density ($n_e$), emission measure (EM) and the in-band spectral index value.}
    \end{table}    
    \end{landscape}


    \begin{table*}[h]
        \centering
        \caption{SiO (5--4) line parameters.}
        \begin{tabular}{c c c c c c c}
        \hline
        ID & Source Name & No. of & Integrated Intensity & Velocity & Width & $\mathrm{T}_{\mathrm{peak}}$ \\
         & & components & (K km\,s$^{-1}$) & (km\,s$^{-1}$) & (km\,s$^{-1}$) & (mK)\\
        \hline
        1& G358.721--0.129 & 5 & 0.077 $\pm$ 0.052 & --82.29 $\pm$ 1.30 & 6.18 $\pm$ 3.10 & 11.628 \\
         & & & 0.168 $\pm$ 0.065 & --72.67 $\pm$ 3.11 & 14.94 $\pm$ 6.26 & 10.566 \\
         & & & 0.119 $\pm$ 0.046 & --46.57 $\pm$ 2.81 & 14.61 $\pm $6.36 & 7.653 \\
         & & & 0.040 $\pm$ 0.023 & --8.98 $\pm$ 1.10 & 3.64 $\pm$ 1.93 & 10.298 \\
         & & & 0.175 $\pm$ 0.041 & 4.59 $\pm$ 1.46 & 12.40 $\pm$ 3.13 & 13.243 \\
         \hline
        2& G358.762--0.131 & 3 & 0.514 $\pm$ 0.193 & --60.68 $\pm$ 2.46 & 14.86 $\pm$ 3.84 & 32.515 \\
         & & & 1.081 $\pm$ 0.254 & --41.28 $\pm$ 1.38 & 18.89 $\pm$ 4.64 & 53.741 \\
         & & & 0.256 $\pm$ 0.098 & --18.94 $\pm$ 2.55 & 14.66 $\pm$ 4.70 & 16.409 \\
           \hline
        3& G358.796--0.102 & 2 & 2.048 $\pm$ 0.038 & --124.03 $\pm$ 0.11 & 12.11 $\pm$ 0.28 & 158.82 \\
         & & & 9.570 $\pm$ 0.076 & --114.34 $\pm$ 0.01 & 21.78 $\pm$ 0.21 & 412.71 \\
           \hline
        4& G358.734--0.116 & 5 & 0.118 $\pm$ 0.022 & --209.85 $\pm$ 0.67 & 5.92 $\pm$ 1.25 & 18.65 \\
         & & & 0.069 $\pm$ 0.030 & --170.81 $\pm$ 1.32 & 6.12 $\pm$ 3.29 & 10.58 \\
         & & & 0.397 $\pm$ 0.048 & --148.48 $\pm$ 1.15 & 18.56 $\pm$ 2.73 & 20.08 \\
         & & & 0.781 $\pm$ 0.059 & --56.55 $\pm$ 0.66 & 19.12 $\pm$ 1.83 & 38.36 \\
         & & & 0.055 $\pm$ 0.028 & --42.87 $\pm$ 0.70 & 3.23 $\pm$ 1.43 & 15.86 \\
           \hline
        5& G358.721--0.109 & 2 & 0.132 $\pm$ 0.039 & --203.1 $\pm$ 1.30 & 8.07 $\pm$ 2.64 & 15.39 \\
         & & & 0.366 $\pm$ 0.050 & --145.9 $\pm$ 0.86 & 12.77 $\pm$ 1.94 & 26.95 \\
         \hline
        7& G358.691--0.121 & 2 & 0.085 $\pm$ 0.132 & --173.7 $\pm$ 11.63 & 17.38 $\pm$ 12.45 & 4.58 \\
         & & & 0.648 $\pm$ 0.140 & --157.0 $\pm$ 1.74 & 18.60 $\pm$ 2.96 & 32.75 \\
         \hline
        8& G358.714--0.141 & 3 & 0.099 $\pm$ 0.025 & --174.66 $\pm$ 0.56 & 4.73 $\pm$ 1.67 & 19.66 \\
         & & & 0.341 $\pm$ 0.095 & 10.32 $\pm$ 3.86 & 25.76 $\pm$ 9.70& 12.42 \\
         & & & 0.385 $\pm$ 0.123 & 9.05 $\pm$ 0.35 & 7.39 $\pm$ 1.46 & 48.87 \\
         \hline
        9& G358.599--0.079 & 3 & 0.983 $\pm$ 0.013 & --151.3 $\pm$ 0.33 & 13.03 $\pm$ 0.58 & 70.88 \\
         & & & 1.670 $\pm$ 0.060 & --148.7 $\pm$ 0.09 & 26.04 $\pm$ 1.66 & 60.24 \\
         & & & 0.418 $\pm$ 0.057 & --121.7 $\pm$ 1.25 & 19.84 $\pm$ 2.54 & 19.79 \\
         \hline
        10& G358.654--0.119 & 3 & 0.103 $\pm$ 0.070 & --137.68 $\pm$ 1.91 & 10.55 $\pm$ 4.32 & 9.21 \\
          & & & 0.999 $\pm$ 0.159 & --122.32 $\pm$ 1.84 & 35.84 $\pm$ 4.35 & 26.19 \\
          & & & 0.140 $\pm$ 0.069 & --100.94 $\pm$ 13.86 & 37.68 $\pm$ 7.76 & 3.50 \\
          \hline
        12& G358.751--0.076 & 5 & 0.076 $\pm$ 0.025 & --179.85 $\pm$ 0.71 & 4.51 $\pm$ 1.88 & 15.82 \\
           & & & 0.209 $\pm$ 0.046 & --119.92 $\pm$ 1.30 & 13.29 $\pm$ 3.78 & 14.75 \\
           & & & 0.108 $\pm$ 0.033 & --104.07 $\pm$ 1.01 & 6.84 $\pm$ 2.09  & 14.81 \\
           & & & 0.115 $\pm$ 0.045 & --87.48 $\pm$ 2.44 & 13.41 $\pm$ 6.65 & 8.02 \\
           & & & 0.238 $\pm$ 0.045 & --59.57 $\pm$ 1.69 & 17.04 $\pm$ 3.53 & 13.12 \\
        \hline
        13& G358.762--0.087 & 4 & 0.089 $\pm$ 0.030 & --149.1 $\pm$ 1.45 & 8.04 $\pm$ 2.81 & 10.44 \\
          & & & 0.435 $\pm$ 0.055 & --112.1 $\pm$ 1.40 & 22.32 $\pm$ 3.15 & 18.30 \\
          & & & 0.177 $\pm$ 0.046 & --84.51 $\pm$ 1.63 & 12.77 $\pm$ 4.0 & 13.06 \\
          & & & 0.146 $\pm$ 0.047 & --56.49 $\pm$ 2.65 & 16.92 $\pm$ 6.21 & 8.00 \\
        \hline
        14& G358.779--0.117 & 6 & 0.639 $\pm$ 0.086 & --134.3 $\pm$ 2.79 & 42.14 $\pm$ 6.58 & 14.24 \\
          & & & 0.104 $\pm$ 0.086 & --74.82 $\pm$ 5.21 & 12.19 $\pm$ 9.97 & 8.03 \\
          & & & 0.212 $\pm$ 0.109 & --62.37 $\pm$ 1.42 & 8.85 $\pm$ 4.08 & 22.49 \\
          & & & 0.491 $\pm$ 0.076 & --48.54 $\pm$ 0.73 & 11.49 $\pm$ 2.16 & 40.11 \\
          & & & 0.086 $\pm$ 0.040 & --37.35 $\pm$ 0.60 & 3.52 $\pm$ 1.40 & 23.06 \\
          & & & 0.488 $\pm$ 0.060 & --25.75 $\pm$ 0.79 & 13.72 $\pm$ 2.13 & 33.42 \\
        \hline
        15& G358.798--0.117 & 4 & 2.247 $\pm$ 0.375 & --135.16 $\pm$ 0.26 & 14.69 $\pm$ 0.91 & 143.69 \\
          & & & 2.728 $\pm$ 0.422 & --123.11 $\pm$ 1.92 & 28.01 $\pm$ 2.11 & 91.47 \\
          & & & 0.636 $\pm$ 0.071 & --53.64 $\pm$ 0.63 & 13.18 $\pm$ 1.76 & 45.32 \\
          & & & 0.863 $\pm$ 0.089 & --23.27 $\pm$ 1.44 & 28.15 $\pm$ 3.47 & 28.80 \\
        \hline
        16& G358.808--0.086 & 5 & 0.169 $\pm$ 0.046 & --119.9 $\pm$ 1.07 & 8.39 $\pm$ 2.95 & 18.96 \\
          & & & 0.082 $\pm$ 0.038 & --110.0 $\pm$ 0.85 & 4.06 $\pm$ 1.82 & 18.96 \\
          & & & 0.248 $\pm$ 0.096 & --87.62 $\pm$ 4.70 & 27.16 $\pm$ 12.77 & 8.57 \\
          & & & 0.080 $\pm$ 0.043 & --68.12 $\pm$ 0.84 & 4.09 $\pm$ 1.93 & 18.46\\
          & & & 0.500 $\pm$ 0.079 & --52.42 $\pm$ 1.06 & 16.54 $\pm$ 3.52 & 28.43\\
        \hline
        17& G358.808--0.131 & 5 & 0.848 $\pm$ 0.134 & --121.71 $\pm$ 0.63 & 11.23 $\pm$ 1.65 & 70.92\\
          & & & 1.190 $\pm$ 0.218 & --97.74 $\pm$ 1.57 & 25.02 $\pm$ 5.311 & 44.68 \\
          & & & 0.347 $\pm$ 0.186 & --46.42 $\pm$ 11.26 & 32.73 $\pm$ 8.56 & 9.97 \\
          & & & 1.676 $\pm$ 0.214 & --42.66 $\pm$ 0.35 & 14.00 $\pm$ 1.12 & 112.45 \\
          & & & 2.140 $\pm$ 0.315 & --37.73 $\pm$ 4.91 & 53.35 $\pm$ 9.71 & 37.69 \\
        \hline    
        \end{tabular}
        
        \label{sio}
    \end{table*}

    \begin{figure*}[h]
    \centering
    \includegraphics[width=\textwidth]{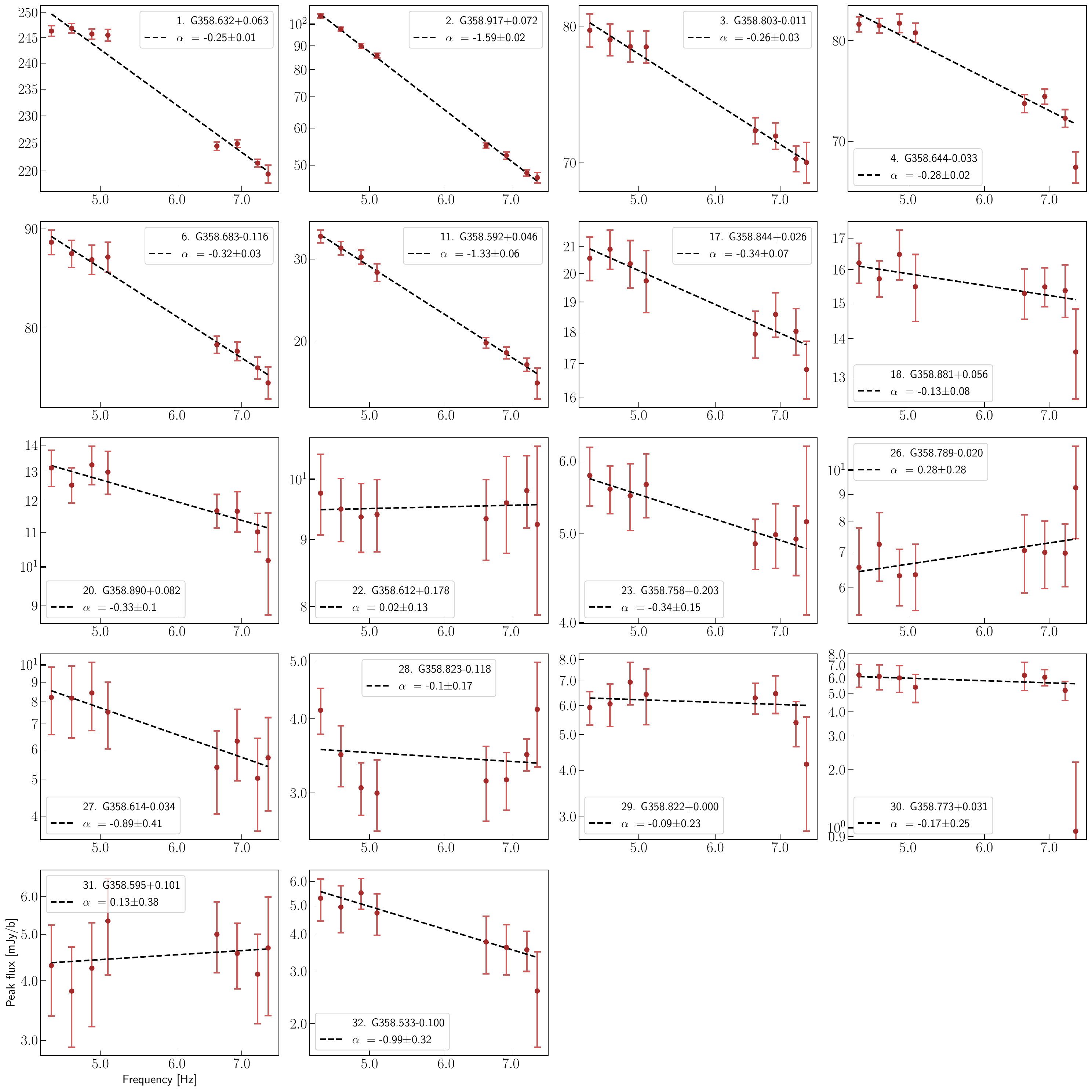}
    \caption{Spectral index plots for the \ion{H}{ii} regions. The spectral index ($\alpha$), source name and ID are given in each panel.}
    \label{si}
    \end{figure*}

    \begin{figure*}[h]
    \centering
    \includegraphics[width=\textwidth]{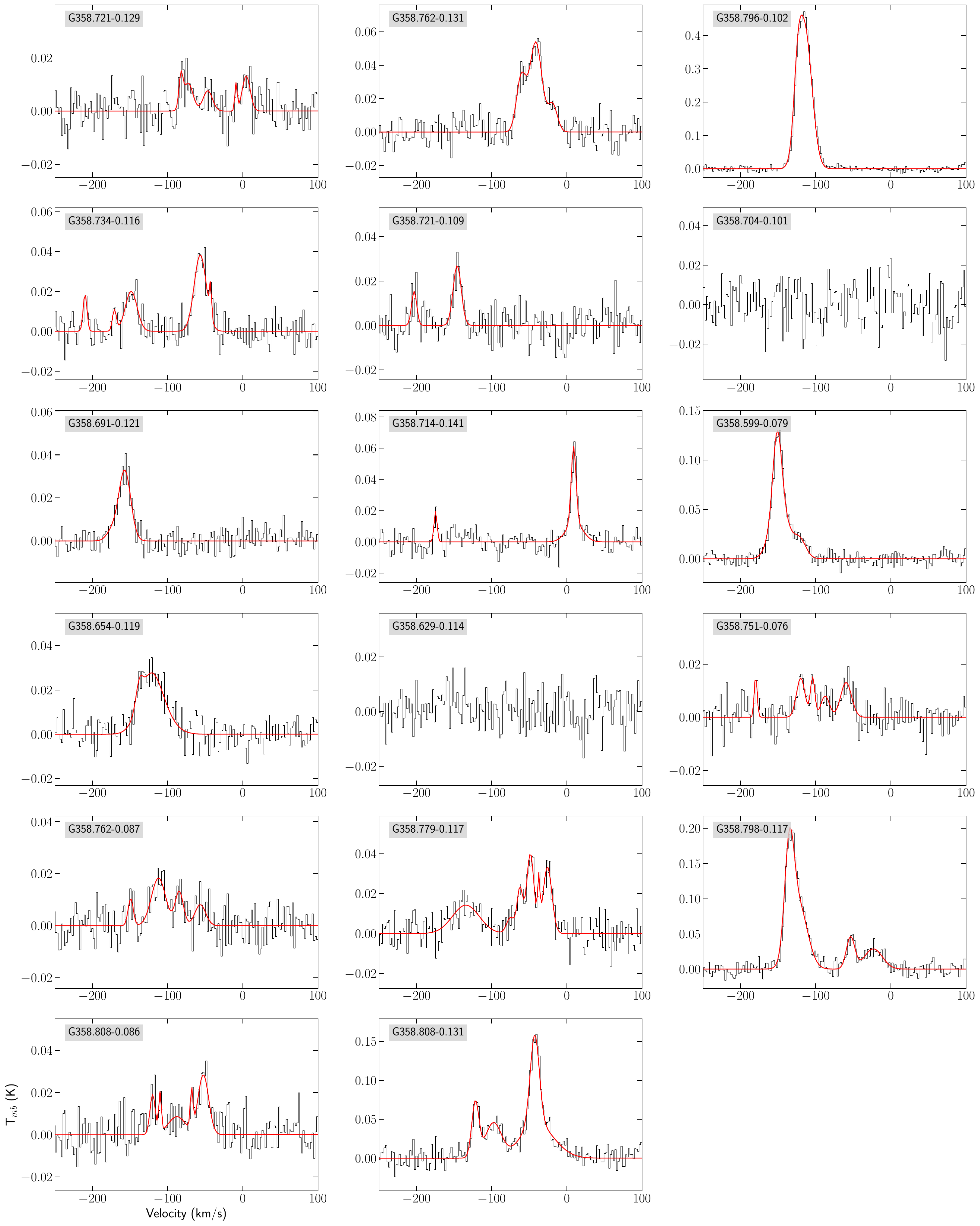}
    \caption{Spectral line (shown in red) of the SiO J=5--4 transition (rest frequency of 217.1050 GHz) for each source on the $T_{\text{mb}}$ scale. The observed spectra are shown in black. The source names are presented in each panel and the plots show a --300 to +100 km\,s$^{-1}$ velocity range.}
    \label{spectrum_of_AGAL_sources}
    \end{figure*}
    
    \end{appendix}
    
\end{document}